\ifdefined\pdfminorversion
\fi
\documentclass[acmsmall,screen]{acmart}
\acmJournal{TOSEM}
\acmDOI{}

\def\JITAtlasCompactEvidence{}
\def\JITAtlasInlineEvidence{}
\usepackage[english]{babel}

\makeatletter
\@ifclassloaded{acmart}{}{\usepackage{amsmath,amssymb}}
\makeatother
\usepackage{graphicx}
\DeclareGraphicsExtensions{.pdf,.png,.jpg}

\usepackage{enumitem}
\usepackage{booktabs}
\usepackage{tabularx}
\usepackage{longtable}
\usepackage{multirow}
\usepackage{makecell}
\usepackage{relsize}
\usepackage{needspace}
\usepackage{placeins}
\usepackage{flafter}
\usepackage[ruled,vlined,linesnumbered]{algorithm2e}

\usepackage{xcolor}
\usepackage{colortbl}
\usepackage[most]{tcolorbox}
\usepackage[normalem]{ulem}

\definecolor{darkgreen}{RGB}{0,100,0}
\definecolor{codegreen}{rgb}{0,0.5,0}
\definecolor{codepurple}{rgb}{0.58,0,0.82}
\definecolor{codegray}{rgb}{0.5,0.5,0.5}
\usepackage{listings}
\lstdefinestyle{mystyle}{
  commentstyle=\color{codegreen},
  keywordstyle=\bfseries,
  stringstyle=\color{codepurple},
  basicstyle=\ttfamily\scriptsize,
  breaklines=true,
  captionpos=b,
  keepspaces=true,
  tabsize=2
}
\lstdefinestyle{casecode}{
  style=mystyle,language=C,
  basicstyle=\ttfamily\fontsize{8}{9.5}\selectfont,
  commentstyle=\color{darkgreen},backgroundcolor=\color{black!1},
  frame=single,rulecolor=\color{black!35},framerule=0.35pt,
  framesep=3pt,numbers=none,aboveskip=0pt,belowskip=0pt,
  xleftmargin=3pt,xrightmargin=3pt,columns=fullflexible
}

\makeatletter
\@ifpackageloaded{hyperref}{}{\usepackage[hidelinks]{hyperref}}
\makeatother

\newif\ifcomments
\commentsfalse
\ifcomments
  \IfFileExists{todonotes.sty}{%
    \usepackage{todonotes}
  }{%
    \newcommand{\todo}[2][]{\textcolor{red}{\textbf{TODO:} #2}}
  }
\else
  \newcommand{\todo}[2][]{}
\fi

\newif\ifinlineevidence
\ifdefined\JITAtlasInlineEvidence
  \inlineevidencetrue
\else
  \inlineevidencefalse
\fi
\newif\ifextendedevidence
\ifdefined\JITAtlasCompactEvidence
  \extendedevidencefalse
\else
  \extendedevidencetrue
\fi

\newcommand{\EvalDesignActionProduce}{61.88}
\newcommand{\EvalDesignActionReplay}{29.80}
\newcommand{\EvalDesignBranchProduce}{20.56}
\newcommand{\EvalDesignBranchReplay}{5.34}
\newcommand{\EvalDesignOutcomeProduce}{1.95}
\newcommand{\EvalDesignOutcomeReplay}{0.73}
\newcommand{\EvalDesignBorrowProduce}{4.64}
\newcommand{\EvalDesignBorrowReplay}{2.15}
\newcommand{\EvalDesignPeakMemory}{65.8}

\newcommand{\papertitle}{{\tech}: Checking Lifetime and Protection Obligations in
Compiler Rewrites}
\newcommand{\papertitleshort}{{\tech}}
\newcommand{\paperkeywords}{Program analysis, compiler validation, object lifetime,
CPython, JIT}

\newcommand{\tech}{\mbox{\textnormal{\textsc{JITAtlas}}}}

\newcommand{\PyJITBench}{\mbox{\textnormal{\textsc{PyJITBench}}}}

\newcommand{\authorAname}{Baihong Chen}
\newcommand{\authorAaffil}{Utah State University}
\newcommand{\authorAcity}{Logan}
\newcommand{\authorAcountry}{USA}
\newcommand{\authorAemail}{b.chen@usu.edu}

\newcommand{\authorCname}{Wen Li}
\newcommand{\authorCaffil}{Utah State University}
\newcommand{\authorCcity}{Logan}
\newcommand{\authorCcountry}{USA}
\newcommand{\authorCemail}{awen.li@usu.edu}

\newif\ifreferences
\referencestrue

\newcommand{\rqfinding}[1]{%
  \par\smallskip
  \noindent\fbox{%
    \begin{minipage}{\dimexpr\linewidth-2\fboxsep-2\fboxrule\relax}
      \small\textbf{Finding.} #1
    \end{minipage}%
  }%
  \par\smallskip
}

\providecommand{\Description}[1]{}

\ifcomments
  
  \newcommand{\wen}[1]{\todo[inline,color=red!40]{Wen: #1}}
\else
  
  \newcommand{\wen}[1]{}
\fi

\definecolor{DarkOrange}{rgb}{0.8,0.3,0.0}
\definecolor{DarkCyan}{rgb}{0.0,0.55,0.55}
\definecolor{backcolour}{rgb}{0.95,0.95,0.92}

\newcommand{\artifacturl}{https://github.com/awen-li/JITAtlas}
\newcommand{\artifact}[1]{\href{\artifacturl}{#1}}

\newcommand{\circleone}[1]{%
  \tikz[baseline=(char.base)]{
    \node[
      shape=circle,
      draw=black,
      fill=white,
      text=black,
      inner sep=1pt
    ] (char) {\small #1};
  }%
}

\makeatletter
\newcommand{\AlgorithmMinipageSetup}[1]{%
  \@twocolumnfalse
  \SetAlFnt{#1}%
  \SetAlCapFnt{#1}%
  \SetAlCapNameFnt{#1}%
}
\makeatother

\makeatletter
\patchcmd{\@secfont}{\MakeTextUppercase}{}{}{}
\makeatother

\setcopyright{none}
\begin{document}

\title[\papertitleshort]{\papertitle}

\author{\authorAname}
\affiliation{%
  \institution{\authorAaffil}
  \city{\authorAcity}
  \country{\authorAcountry}
}
\email{\authorAemail}

\author{\authorCname}
\affiliation{%
  \institution{\authorCaffil}
  \city{\authorCcity}
  \country{\authorCcountry}
}
\email{\authorCemail}


\begin{abstract}
Compiler rewrites can preserve returned values while violating object lifetime
or ownership. Such failures arise when a rewrite changes when an object remains
live, whether a borrowed pointer stays supported, or what ownership reaches the
caller. We present \tech{}, a source-level checker that reconstructs reference
and target computations from optimizer code and checks valid uses, continuous
borrow support, and caller-visible ownership.
On a 58-input benchmark constructed and labeled independently of \tech{} output,
\tech{} detects all 29 defects and reports none of 29 safe inputs; PyRefcon
detects 20 defects and CpyChecker 18. Across three CPython snapshots, \tech{}
extracts 3,332 rewrite alternatives, retains 801 under recovered receiving
contexts, and reports 194 violations. Manual review confirms 92 reports, giving
47.42\% precision. End-to-end analysis takes 35.56--42.31~s per snapshot.
\tech{} further identifies 15 new bugs across five failure classes, each with
a reproducible proof of concept.
\end{abstract}

\maketitle

\section{Introduction}\label{sec:intro}

A compiler rewrite replaces one computation with another. The original
computation is the \emph{reference}, and the rewritten computation is the
\emph{target}. Rewrites can remove reference-count operations, replace lookups
with cached pointers, eliminate runtime checks (guards), or reorder cleanup
around an operation. These changes rely on facts established during
optimization, such as an object's type, value location, or owner that keeps it
alive~\cite{cinderx-refcount,cpython-jit-rq1-doc}.
Preserving the returned value alone is therefore insufficient. A rewrite can
return the same pointer while changing when it is safe to use, what keeps its
object alive, or which runtime fact justifies an optimized access. A pointer
can retain the same address after its object has been freed. The type, stored
value, or ownership relation assumed by an optimization can also change while
the object remains live. A correct rewrite must therefore preserve object
lifetime and the type, slot, and ownership conditions at each use.

A CPython attribute rewrite illustrates the problem. A borrowed pointer does
not own the object to which it refers. The reference acquires ownership of a
borrowed attribute pointer before releasing the lookup receiver. The target
reverses these actions. Releasing the receiver can invoke a finalizer, a
callback during object destruction, that replaces the attribute and releases
the payload's last owner. The target then acquires ownership through a stale
pointer. Final reference counts can still balance even though the object became
invalid before acquisition.
A checker for such rewrites must recover the executed operations, track
lifetime through callbacks and mutations, and preserve the ownership, exits,
and visible effects required by the caller.

Existing techniques address parts of this problem. Translation validation
compares reference and target programs once both programs and their semantics
are available~\cite{pnueli-translation-validation,alive2}. Optimizer source,
however, does not provide this pair directly. The selected computations,
their entry conditions, and the effects of runtime helpers must first be
recovered from the compiler implementation~\cite{cpython-stackrefs-rq1-doc,cpython-jit-rq1-doc}.
Ownership-aware representations and reference-count analyses track borrowing,
transfer, and release~\cite{swift-ossa,perceus,pungi,pyrefcon}. Rewrite
checking must also determine whether those actions remain valid when their
order changes, callbacks mutate supporting state, and the rewritten
computation transfers results back to its caller.

The problem raises three challenges. \emph{C1: Recover the selected
computation.} Optimizer source does not directly expose a reference--target
pair. The checker must reconstruct it from the rewrite conditions, helper
calls, emitted-operation-buffer writes, and the code that supplies inputs and
receives results.
\emph{C2: Check protection through intervening effects.} A fact that holds
at entry can become false before use. A borrowed object can also lose the
owner or runtime root that keeps it alive even if support exists at both
endpoints. The checker must bind each borrow to its original object and record
any support loss, even if support later returns.
\emph{C3: Preserve the caller's contract.} The reference and target must
agree on the identity and ownership of returned objects, the exit taken, and
the effects visible to the caller.

We present \tech{}, a source-level checker for lifetime and protection
obligations in compiler rewrites. A runtime contract specifies an operation's
requirements, effects, and possible outcomes. \tech{} addresses the three
challenges in three stages. \emph{Source recovery} reconstructs the ordered
reference and target computations and binds their actions to runtime contracts
(C1). \emph{Lifetime checking} symbolically executes these actions to check
valid uses, continuous borrow support, and ownership at exits (C2).
\emph{Reference comparison} derives caller obligations from the reference and
checks target outcomes against reference executions possible on the same
inputs (C3). The production adapter targets CPython's just-in-time (JIT)
compiler.
Two design principles shape these checks. First, a fact proposed by the
optimizer must be proved rather than assumed at entry. A deleted guard
therefore cannot justify its own removal. Second, borrow support is an interval
property. If support disappears before a use, restoring it later does not
repair the earlier gap. Requirements derived from the reference expose missing
ownership at exits. Matching reference executions determines whether the
target's outcomes are permitted by the reference.

The evaluation uses four populations for different purposes. An independently
constructed and labeled 58-input C benchmark measures controlled accuracy
against PyRefcon and CpyChecker; it was subsequently used during
implementation. Additional development controls test individual mechanisms.
Production CPython rewrites measure report precision and the coverage of the
available source and runtime semantics. Frozen transfer studies test the fixed
checker and contracts on separately selected revisions.
On the shared accuracy benchmark, \tech{} detects all 29 labeled defects and
reports none of 29 safe inputs; PyRefcon detects 20 defects and CpyChecker 18.
Across three CPython snapshots, production analysis extracts 3,332 alternatives
and retains 801 under recovered receiving contexts. It reports 194 violations;
manual review confirms 92, giving 47.42\% precision. Ablations show that
endpoint-only borrow checking misses four violations, recovering caller-visible
ownership exposes one additional violation, and precise receiving-context
recovery removes 92 false reports without removing a true report. \tech{}
identifies 15 new bugs across five failure classes, each with a reproducible
proof of concept.
This paper makes three \textbf{contributions}:
\begin{itemize}
  \item a lifetime-preservation property for compiler rewrites that separates
  valid uses, continuous borrow support, and caller-visible ownership and
  effects;
  \item a source-level checking method that reconstructs selected rewrites,
  tracks lifetime through ordered runtime effects, and derives caller
  obligations from the reference without assuming optimizer-proposed facts;
  and
  \item an implementation and CPython evaluation that measures the contribution
  of these checks and identifies 15 new bugs in five failure classes, each with
  a reproducible proof of concept.
\end{itemize}

\vspace{3pt}
\noindent\textbf{\underline{Open science.}}
We release the implementation, reproduction scripts and outputs, and
\PyJITBench{}, which packages the controlled inputs and experiment records,
at \artifact{{\tech}}.

\section{Background and Motivation}\label{sec:background}

\subsection{Objects and Rewrites in CPython}\label{sec:cpython-background}

\begin{figure}[htp]
\centering
\includegraphics[width=\columnwidth]{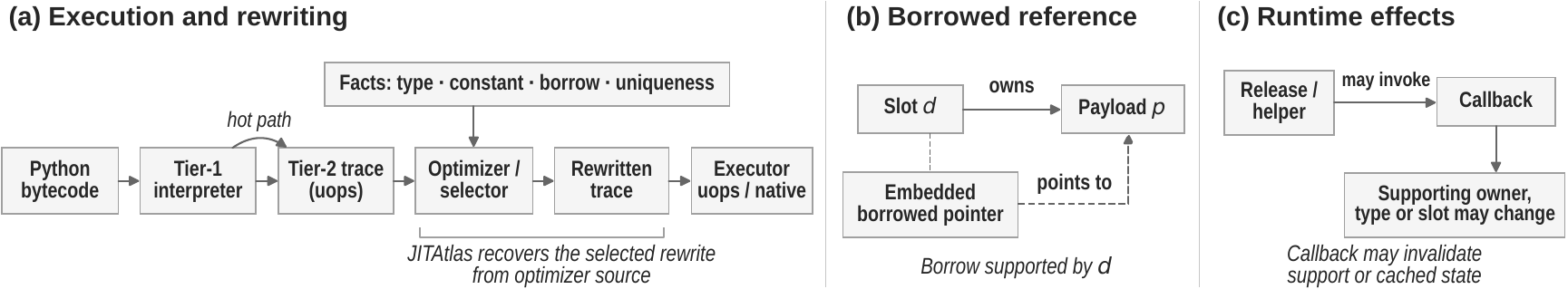}
\caption{CPython execution and rewrite context. \tech{} recovers selected
rewrites from optimizer source. An embedded pointer can borrow an object
supported by an owning runtime location, while callbacks can change the
ownership or runtime state on which later uses depend.}
\Description{Three horizontal panels show execution and rewriting, borrow
support, and runtime effects. {\tech} recovers the selected rewrite from
optimizer source. Slot d owns payload p, supporting an embedded borrowed
pointer to p. A release or helper may invoke a callback that changes the
supporting owner, type, or slot and can invalidate borrow support or cached
state.}
\label{fig:cpython-pipeline}
\end{figure}

Figure~\ref{fig:cpython-pipeline}(a) shows where rewriting occurs in CPython's
execution pipeline. The Tier-1 interpreter executes Python bytecode, while
Tier-2 records hot paths as \emph{traces} of micro-operations (uops). The
optimizer rewrites these traces. Its \emph{selector} uses facts such as an
object's type, constant value, ownership, or uniqueness (having a single
owner) to choose a rewrite. An \emph{executor} stores the rewritten trace and
runs it as uops or generated native code~\cite{python-pep744,cpython-jit-315}.
\tech{} recovers these selected rewrites from optimizer source.

Specialized traces can retain pointers whose validity depends on runtime
ownership, as Figure~\ref{fig:cpython-pipeline}(b) illustrates. Most CPython
objects use reference counting, with cyclic garbage collection for unreachable
cycles~\cite{cpython-refcount-315,cpython-gc-315}. A \emph{handle} is a runtime
reference that identifies an object and records its ownership state. An owned
handle keeps the object alive until its ownership is released or transferred.
A borrowed pointer does not keep the object alive by itself; it depends on
another owner or runtime root, such as a reference held by the
interpreter~\cite{cpython-refcount-315,cpython-stackrefs-rq1-doc}. In the
figure, slot $d$ owns payload $p$ and keeps it alive for the embedded pointer
that borrows it. Immortal objects are a special case: they remain live
independently of their reference count~\cite{python-pep683}.

This support can change during execution, as
Figure~\ref{fig:cpython-pipeline}(c) shows. Releasing an object or calling a
runtime helper can invoke Python code, including a finalizer during object
destruction~\cite{cpython-refcount-315}. Such callbacks can remove an owner,
replace stored values, or change a type's dispatch slots, which determine the
operations that execute. These changes can invalidate a later use of cached
state. Runtime guards check assumptions on the current path, type watchers
observe changes to registered types~\cite{cpython-type-watchers-315}, and
executor dependencies record the types on which a trace
depends~\cite{cpython-optimizer-315}. These mechanisms protect cached type and
slot assumptions.

\subsection{Motivation Example}
\label{sec:running-example}\label{sec:threat-model}

Consider an attribute rewrite that changes the order of two ownership actions.
The receiver object $r$ has a class attribute whose slot $d$ owns payload $p$.
Both the reference and target borrow a pointer to $p$ from $d$. The reference
acquires ownership of $p$ before releasing $r$, whereas the target releases
$r$ first (Figure~\ref{fig:attribute-running}).
Suppose $d$ is the only owner of $p$ and releasing $r$ destroys the receiver.
Its finalizer replaces the attribute stored in $d$, releasing the old payload.
In the target, this removes $p$'s last owner before \path{acquire(p)}. The
borrowed pointer still contains the old address, but $p$ is no longer live, so
the acquisition uses a stale pointer. In the reference, the earlier
acquisition creates another owner and keeps $p$ live through the same callback.
Acquiring ownership afterward cannot repair the earlier lifetime gap.
In CPython, the target trace realizes this order as
\path{POP_TOP; _LOAD_CONST_INLINE}. \path{POP_TOP} releases the receiver, and
\path{_LOAD_CONST_INLINE} then acquires ownership through the embedded payload
pointer.

\begin{figure}[!htbp]
\centering
\lstset{language=C,basicstyle=\ttfamily\fontsize{7.5}{8.8}\selectfont,
  commentstyle=\color{darkgreen},backgroundcolor=\color{black!1},
  frame=single,rulecolor=\color{black!35},framerule=0.35pt,
  framesep=3pt,numbers=none,aboveskip=0pt,belowskip=0pt,
  xleftmargin=3pt,xrightmargin=3pt,columns=fullflexible}
\begin{minipage}[t]{0.49\linewidth}
\begin{lstlisting}
// Reference
p = borrow(d);
acquire(p);
release(r);      // may invoke callback
return_owned(p);
\end{lstlisting}
\end{minipage}\hfill
\begin{minipage}[t]{0.49\linewidth}
\begin{lstlisting}
// Target
p = borrow(d);
release(r);      // callback may free p
acquire(p);      // invalid if p is freed
return_owned(p);
\end{lstlisting}
\end{minipage}
\caption{Semantic actions in the attribute rewrite. Slot $d$ is initially
$p$'s only owner. The reference acquires another owner before releasing $r$;
the target releases $r$ first, so a callback can free $p$ before acquisition.}
\Description{Two side-by-side code boxes compare the reference and target.
Both borrow payload p from slot d. The reference acquires p before releasing
receiver r. The target releases r first, allowing a callback to free p before
the later acquisition.}
\label{fig:attribute-running}
\end{figure}

The analysis must connect the borrowed $p$ to its supporting slot $d$, connect
the release of $r$ to the callback that can remove this support, and preserve
the order of these effects. A check that omits the callback misses the loss of
support between release and acquisition.
The rewrite also has an exit obligation. Its caller expects ownership of the
original payload $p$. Returning a live borrow would avoid the stale
acquisition but still violate this interface. Correctness therefore requires
both lifetime safety during execution and preservation of the ownership and
effects visible to the caller.
These observations lead to three analysis tasks. \emph{Source recovery}
reconstructs the borrow, release, callback, and acquisition in execution order
(C1). \emph{Lifetime checking} follows the captured $p$ and its support through
these effects (C2). \emph{Reference comparison} recovers the owned-result
requirement and checks caller-visible outputs, exits, and effects (C3).

\subsection{Motivating Transformation and Failure Patterns}
\label{sec:empirical-basis}

\begin{figure}[!ht]
  \centering
  \includegraphics[width=0.8\columnwidth]{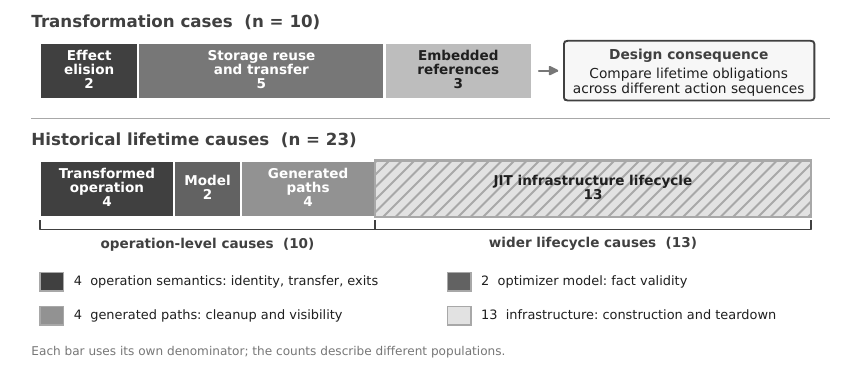}
  \vspace{-5pt}
  \caption{Motivating transformation and failure patterns. Ten transformation
  cases group lifetime-sensitive rewrites by mechanism; 23 historical causes
  group failures by JIT component. The two illustrative collections are
  separate.}
  \Description{Two grayscale stacked bars summarize separate illustrative
  collections. The first divides ten transformation cases into two effect
  eliminations, five storage-reuse or ownership-transfer cases, and three
  embedded-reference cases. The second divides 23 historical causes into four
  transformed-operation causes, two optimizer-model causes, four generated
  cleanup or visibility causes, and 13 JIT-infrastructure lifecycle causes.}
  \label{fig:empirical-background}
\end{figure}

The attribute rewrite is one instance of a broader class of lifetime-sensitive
rewrites. Figure~\ref{fig:empirical-background} summarizes two illustrative
collections: ten transformation cases and 23 historical JIT failure causes.
These cases show the mechanisms that the analysis must represent; the counts
describe the included cases, not their prevalence in CPython.
The transformation cases group into effect elimination, storage reuse or
ownership transfer, and embedded references. The historical causes group into
transformed operations, optimizer models, generated cleanup or visibility,
and JIT lifecycle code. Upstream repairs include a use-after-free in optimized
Unicode concatenation~\cite{cpython-fix-unicode-append-uaf}, treating reference
decrements as escaping operations~\cite{cpython-fix-dealloc-escape}, and an
executor-creation leak~\cite{cpython-fix-executor-owner-leak}.
The transformed-operation, optimizer-model, and generated-path cases expose
the same obligations as the attribute rewrite: each use must be valid when it
occurs, each borrow must remain supported throughout its interval, and each
exit must satisfy the ownership and effects required by the caller.

\section{Problem Definition}\label{sec:problem}

A rewrite is correct only if it preserves the lifetime conditions of the
objects it uses. Three conditions matter: every use must operate on a valid
object, every borrowed object must remain supported while the borrow is active,
and every exit must satisfy the ownership required by the caller.

\vspace{3pt}
\noindent
\circleone{1} \textbf{Transformation model.}
Let $S_R$ be the reference computation and $S_T$ the rewritten target.
The optimizer condition $\phi$ selects the rewrite, $\rho$ relates their
inputs and observable results, and the runtime contracts $\Gamma$ define
operation requirements, effects, and outcomes. We model the transformation as
\[
  \tau=\langle S_R,\phi,S_T,\rho,\Gamma\rangle.
\]
Let $P_0$ denote the premises available from the caller at entry.
For a computation $S$ and initial state $s$,
$\operatorname{Exec}_{\Gamma}(S,s)$ is the set of executions allowed by
$\Gamma$. An execution is a sequence of states $\sigma_i$ connected by actions
$a_i$:
\[
  \pi=\sigma_0\xrightarrow{a_0}\sigma_1
      \xrightarrow{a_1}\cdots.
\]
The actions include intermediate effects, such as callbacks triggered when an
object is released.

\vspace{3pt}
\noindent
\circleone{2} \textbf{Lifetime safety.}
Lifetime safety consists of three obligations.
\emph{Point obligation.}
At a use $u$ in state $\sigma_i$, the handle must be valid, its object must be
live, and the runtime facts required by the operation must hold. We write this
condition as $\operatorname{Req}_{\Gamma}(u,\sigma_i)$. It is checked before
the action takes effect.
\emph{Interval obligation.}
A borrow $b$ records the object $o_b$ to which it refers. While $b$ is active,
$o_b$ must remain supported by an owner or runtime root. We write this
condition as $\operatorname{Sup}_{\Gamma}(b,o_b,\sigma_i)$. The supporting
owner or root may change, but support cannot disappear and the borrowed object
cannot change.
\emph{Exit obligation.}
At exit $x$, ownership must be released, transferred, or exported as required
by the caller. We write this condition in state $\sigma_i$ as
$\operatorname{Exit}_{\Gamma}(x,\sigma_i)$.

Let $\operatorname{Uses}(\pi)$ be the uses in $\pi$ with their positions,
$\operatorname{Active}_b(\pi)$ the positions where borrow $b$ is active, and
$\operatorname{Exits}(\pi)$ the exits with their positions. An execution is
safe exactly when all three obligations hold:
\[
\begin{aligned}
 \operatorname{Safe}_{\Gamma}(\pi)\equiv
 &\bigwedge_{(u,i)\in\operatorname{Uses}(\pi)}
       \operatorname{Req}_{\Gamma}(u,\sigma_i)\\
 {}\land\;&\bigwedge_b\bigwedge_{i\in\operatorname{Active}_b(\pi)}
       \operatorname{Sup}_{\Gamma}(b,o_b,\sigma_i)\\
 {}\land\;&\bigwedge_{(x,i)\in\operatorname{Exits}(\pi)}
       \operatorname{Exit}_{\Gamma}(x,\sigma_i).
\end{aligned}
\]

\vspace{3pt}
\noindent
\circleone{3} \textbf{Relational preservation.}
A safe target execution must also produce an outcome permitted by the
reference. Let $\operatorname{Obs}_{\Gamma}(\pi)$ contain the caller-visible
outputs, ownership, aliases, exits, and ordered effects of execution $\pi$.
The relation $\sim_{\rho}$ states when two such observations correspond, and
$\pi\downarrow$ means that $\pi$ terminates.
For every pair of entry states $(s_R,s_T)$ related by $\rho$ and satisfying
$P_0$, and every target execution
$\pi_T\in\operatorname{Exec}_{\Gamma}(S_T,s_T)$, we require
\[
\begin{aligned}
 &\operatorname{Safe}_{\Gamma}(\pi_T)\ \land\\
 &\left(\pi_T\downarrow\Longrightarrow
   \exists\pi_R\in\operatorname{Exec}_{\Gamma}(S_R,s_R):
   \pi_R\downarrow\land\operatorname{Safe}_{\Gamma}(\pi_R)\land
   \operatorname{Obs}_{\Gamma}(\pi_R)
   \sim_{\rho}\operatorname{Obs}_{\Gamma}(\pi_T)\right).
\end{aligned}
\]
The reference and target may use different internal ownership actions, but
the target's caller-visible outcome must match an outcome produced by a
realizable reference execution.

\vspace{3pt}
\noindent
\circleone{4} \textbf{Optimizer proposals.}
The selecting condition $\phi$ may propose facts used to justify the rewrite.
Let $\Delta$ contain these proposed facts, and let $\Phi_u$ contain the facts
established by executed guards and actions before use $u$. For every proposed
fact $F\in\Delta$ required at $u$, correctness requires
\[
  P_0\land\Phi_u \models F.
\]
The proof cannot therefore assume $\Delta$. In particular, removing a guard
cannot be justified by assuming the fact that the guard was meant to
establish.

\section{{\tech} Design}\label{sec:design}

\begin{figure*}[htp]
  \centering
  \includegraphics[width=\textwidth]{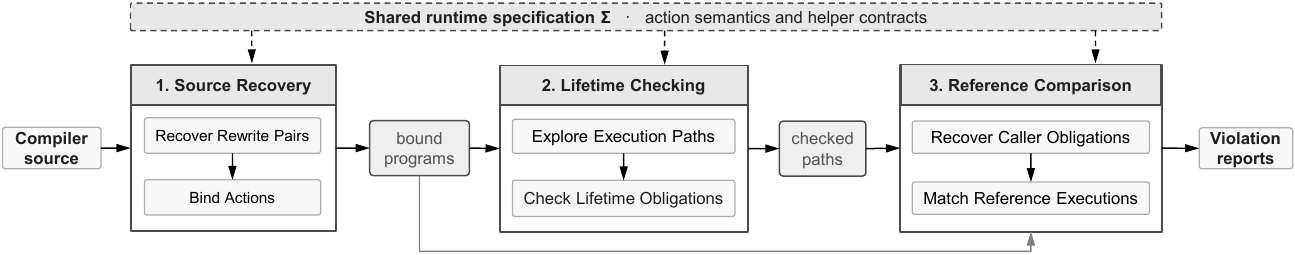}
  \caption{Overview of \tech{}. Source recovery reconstructs and binds the
  selected rewrite, lifetime checking checks its obligations in execution
  order, and reference comparison recovers caller obligations and matches
  reference executions.}
  \Description{Three panels show source recovery, lifetime checking, and
  reference comparison. Compiler source is recovered into bound programs,
  lifetime checking explores their execution paths and checks lifetime
  obligations, and reference comparison recovers caller obligations and
  matches reference executions.}
  \label{fig:technique-overview}
\end{figure*}

\tech{} checks each rewrite using an executable runtime specification $\Sigma$.
The specification realizes the runtime contracts $\Gamma$ from
Section~\ref{sec:problem}, including operation preconditions, ordered effects,
and possible outcomes. A \emph{source binding} maps a recovered operation and
its operands to a runtime contract. Applying these bindings produces a
\emph{bound program} that the checker can execute symbolically. If an action
cannot be fully bound, it remains \emph{opaque}, and any check that depends on
that action remains incomplete.
Figure~\ref{fig:technique-overview} shows the three stages of the analysis.
\emph{Source recovery} reconstructs the selected reference and target
computations from compiler source and binds their actions (C1).
\emph{Lifetime checking} executes the bound programs in order and checks point,
interval, and exit obligations (C2). \emph{Reference comparison} derives
caller obligations from the reference and matches the target's outputs, exits,
and visible effects against realizable reference executions (C3).

\subsection{Source Recovery}\label{sec:dfg}

Source recovery turns compiler source into bound reference and target programs.
As Figure~\ref{fig:technique-overview} shows, it has two steps: recover the
selected reference--target pairs and bind their actions to runtime contracts.
The recovered programs retain the executed actions, their order and operands,
the code that supplies their inputs and receives their results, and the runtime
meaning of each bound action.

\noindent\textbf{Recover rewrite pairs.}
{\tech} first inventories the source actions that may participate in a rewrite;
let $I$ denote this inventory. It then follows helper calls and
operation-buffer writes to reconstruct the reference and target selected on
each optimizer path. Buffer writes are applied in path order, and each pair
retains its selection condition. If several paths produce the same pair, their
conditions are combined.
{\tech} also recovers the executed code that supplies the rewrite's inputs and
receives each exit. It composes this code with the selected operations to form
reference and target regions $R$ and $T$, and relates their inputs and exits by
$\rho$. If the input or exit requirements of $R$ and $T$ cannot be related by
$\rho$, {\tech} records a construction failure and forms no comparison.

For CPython, the default configuration recovers the \emph{canonical receiving
contexts}: enclosing bytecode macros and stack facts established by operations
that precede the selected rewrite. An alternative is excluded only when its
selection condition contradicts every recovered context. If no context is
recovered, the alternative remains unconstrained. These facts describe the
state at entry to the selected operation; they do not turn optimizer proposals
$\Delta$ into assumptions.

\noindent\textbf{Bind actions.}
{\tech} matches each recovered action and its operands to a runtime contract
in $\Sigma$. Binding produces programs $L_R$ and $L_T$ whose actions have
executable preconditions, ordered effects, and possible outcomes. It also
produces a binding record $\Lambda$ that identifies source actions with
complete bindings. An action without a complete binding remains opaque.
Binding also separates caller premises $P_0$ from optimizer proposals
$\Delta$. Caller premises are facts available when execution enters the
rewrite. Optimizer proposals are facts used to justify selecting the target;
they are recorded but not assumed at entry.

\begin{figure}[htp]
\centering
\ifdefined\Description
\Description{Source recovery reconstructs ordered reference--target rewrite
pairs from compiler source, attaches their input and exit contexts, binds
recovered actions to runtime contracts, and records caller premises and
optimizer proposals.}
\fi
\begin{minipage}{0.7\linewidth}
\AlgorithmMinipageSetup{\footnotesize}
\begin{algorithm}[H]
\setlength{\algomargin}{0.35em}
\DontPrintSemicolon
\LinesNumbered
\caption{Recover rewrite pairs and bind actions}
\label{alg:source-protocol}

\SetKwFunction{Inventory}{InventorySourceActions}
\SetKwFunction{Recover}{RecoverOrderedPairs}
\SetKwFunction{Compose}{ComposePrefixAndReceivers}
\SetKwFunction{Compatible}{CompatibleInterfaces}
\SetKwFunction{Failure}{RecordConstructionFailure}
\SetKwFunction{Bind}{BindActions}
\SetKwFunction{Entry}{BindEntryClauses}
\SetKw{Continue}{continue}
\SetKw{Return}{return}

\KwIn{$P$: compiler source; $\Sigma$: runtime specification}
\KwOut{$\mathcal B$: bound comparisons; $K$: construction log}

$I\leftarrow\Inventory(P)$
  \nllabel{ln:recover-inventory}\;
$(\mathcal P,K)\leftarrow\Recover(P)$
  \nllabel{ln:recover-pairs}\;
$\mathcal B\leftarrow\emptyset$\;

\ForEach{$\tau\in\mathcal P$}{
  $(R,T,\rho)\leftarrow\Compose(\tau)$
    \nllabel{ln:recover-compose}\;

  \If{$\neg\Compatible(R,T,\rho)$}{
    $K\leftarrow\Failure(K,\tau)$
      \nllabel{ln:recover-interface}\;
    \Continue\;
  }

  $(L_R,L_T,\Lambda)\leftarrow\Bind(R,T,I,\Sigma)$
    \nllabel{ln:recover-bind}\;
  $(P_0,\Delta)\leftarrow\Entry(\tau,\rho,\Sigma)$
    \nllabel{ln:recover-entry}\;

  $\mathcal B\leftarrow\mathcal B\cup
    \{(L_R,L_T,\rho,P_0,\Delta,\Lambda)\}$
    \nllabel{ln:recover-output}\;
}

\Return{$(\mathcal B,K)$}
  \nllabel{ln:recover-return}\;
\end{algorithm}
\end{minipage}
\end{figure}

Algorithm~\ref{alg:source-protocol} implements these two steps.
$\mathsf{InventorySourceActions}$ records the relevant source actions and their
source identities (line~\ref{ln:recover-inventory}).
$\mathsf{RecoverOrderedPairs}$ follows optimizer paths and applies
operation-buffer writes in order to recover the selected reference--target
pairs and their conditions (line~\ref{ln:recover-pairs}).
For each pair, $\mathsf{ComposePrefixAndReceivers}$ attaches the executed
input-producing and exit-receiving code, producing $R$, $T$, and their
correspondence $\rho$ (line~\ref{ln:recover-compose}).
$\mathsf{CompatibleInterfaces}$ checks whether their input and exit
requirements can be related by $\rho$. If not,
$\mathsf{RecordConstructionFailure}$ records the failed construction
(line~\ref{ln:recover-interface}).
For a compatible pair, $\mathsf{BindActions}$ maps the recovered actions and
operands to contracts in $\Sigma$, producing $L_R$, $L_T$, and $\Lambda$
(line~\ref{ln:recover-bind}). $\mathsf{BindEntryClauses}$ separates the entry
facts into caller premises $P_0$ and optimizer proposals $\Delta$
(line~\ref{ln:recover-entry}). The resulting tuple
$(L_R,L_T,\rho,P_0,\Delta,\Lambda)$ forms one bound comparison
(line~\ref{ln:recover-output}).

The procedure preserves the information required by the obligations in
Section~\ref{sec:problem}. Action order determines the intermediate states in
which uses occur: the same release and acquisition can be safe in one order
and unsafe in another. Input-producing code determines the ownership, aliases,
and guards available at entry. Exit-receiving code determines the ownership
and cleanup required at each exit. The correspondence $\rho$ identifies the
reference and target inputs and exits that must agree.
Binding supplies the runtime meaning of the recovered actions. Compiler source
identifies which actions execute, but does not by itself define their lifetime
preconditions or state changes. Contracts in $\Sigma$ provide these
preconditions, ordered effects, and outcomes. 
Separating $P_0$ from $\Delta$ prevents an optimizer proposal from being
assumed before an executed guard or action establishes it.
The resulting bound comparison contains the information needed to evaluate the
property in Section~\ref{sec:problem}: ordered state changes for point and
interval obligations, exit interfaces for exit obligations, and $\rho$ for
relating caller-visible results. A missing action, binding, or interface
relation leaves the affected check incomplete rather than contributing to a
completed result.

\noindent\textbf{Running example.}
For the attribute rewrite in Figure~\ref{fig:attribute-running}, recovery
reconstructs
\[
\begin{array}{ll}
\textit{reference:} &
\mathsf{borrow}(d);\;
\mathsf{acquire}(p);\;
\mathsf{release}(r);\;
\mathsf{return\_owned}(p),\\[2pt]
\textit{target:} &
\mathsf{borrow}(d);\;
\mathsf{release}(r);\;
\mathsf{acquire}(p);\;
\mathsf{return\_owned}(p).
\end{array}
\]
Recovery preserves the relation between receiver $r$, owning slot $d$, and
captured payload $p$. Binding the release of $r$ adds its possible callback
effect: the callback can replace the value in $d$ and release the old payload.
The bound target therefore contains a path in which the callback occurs after
$\mathsf{borrow}(d)$ and before $\mathsf{acquire}(p)$; the reference acquires
$p$ before the callback can occur.
For this rewrite, source recovery produces $L_R$ and $L_T$, their input/exit
correspondence $\rho$, caller premises $P_0$, optimizer proposals $\Delta$,
and binding record $\Lambda$. In $L_T$, the recovered order and callback
effect expose the intermediate state in which $p$ can lose its last supporting
owner before $\mathsf{acquire}(p)$.

\subsection{Lifetime Checking}\label{sec:detection}

Lifetime checking uses symbolic execution~\cite{king-symbolic-execution} to
explore each bound program and record its lifetime obligations along execution
paths. As Figure~\ref{fig:technique-overview} shows, it has two steps: explore
execution paths and check lifetime obligations. For a bound program $L$, caller
premises $P_0$, and runtime specification $\Sigma$, the result is a set of
checked path records $\mathcal E$.

\noindent\textbf{Explore execution paths.}
{\tech} starts from the entry state established by $P_0$ and executes the
actions in $L$ in program order. A worklist $W$ stores pending path states.
When an action has several possible outcomes, {\tech} creates a successor for
each feasible or undecided outcome and records its path condition and symbolic
state. Terminal paths are added to $\mathcal E$.
Each path record retains the executed actions, path condition, symbolic state,
completed checks, and unresolved effects. Branches and runtime outcomes remain
separate, preserving the conditions and action order under which each
obligation is checked.

\noindent\textbf{Check lifetime obligations.}
{\tech} checks each requirement in the state where it must hold. Point
requirements are checked before an action takes effect. The action is then
applied, and each active borrow whose support may have changed is checked in
the resulting state. Borrow creation, use, transfer, and consumption update
the borrow record while preserving the object captured when the borrow began.
A path that reaches an exit also records its exit requirement.
Required facts must follow from executed code. Let $F$ be a required fact with
predicate $\psi_F$. Operand binding $\mu$ maps the predicate's operands to
program values, symbolic store $\sigma$ records their current state, and
$\chi$ contains $P_0$ and path conditions established by executed code.
{\tech} establishes $F$ at a use only if
\[
  \chi \land \neg\psi_F[\sigma,\mu]
  \quad\text{is unsatisfiable}.
\]
Optimizer proposals $\Delta$ are excluded from $\chi$. A later guard or state
update can establish a fact for a later use, but cannot change an earlier
check.

{\tech} checks continuous support after every action that may affect an active
borrow. Each support predicate records the locations it reads, called its
\emph{read footprint}. If an action writes one of these locations, {\tech}
reevaluates support in the resulting state and updates the footprint. For
example, if support moves from $D[a]$ to $D[b]$, later writes to $D[b]$ must
be checked. A \emph{frame condition} identifies locations left unchanged by an
action, so writes to disjoint locations preserve support, following local
reasoning over disjoint state~\cite{reynolds-separation-logic}. A transferred
borrow retains its captured object, support bindings, and previous checks, so
its active interval continues across the transfer.

\begin{figure}[htp]
\centering
\ifdefined\Description
\Description{Lifetime checking symbolically explores paths of a bound program,
checks point requirements before actions, checks borrow support after relevant
effects, and records completed path states.}
\fi
\begin{minipage}{0.7\linewidth}
\AlgorithmMinipageSetup{\footnotesize}
\begin{algorithm}[H]
\setlength{\algomargin}{0.35em}
\DontPrintSemicolon
\LinesNumbered
\caption{Explore execution paths and check lifetime obligations}
\label{alg:check-lifetime}

\SetKwFunction{Initial}{InitialExecution}
\SetKwFunction{Next}{NextTransition}
\SetKwFunction{Precheck}{AppendPrechecks}
\SetKwFunction{Apply}{ApplyAndCheckSupport}
\SetKwFunction{Borrow}{CheckBorrowTransition}
\SetKwFunction{Terminal}{Terminal}
\SetKw{Return}{return}

\KwIn{$L$: bound program; $P_0$: caller premises;
$\Sigma$: runtime specification}
\KwOut{$\mathcal E$: checked path records}

$W\leftarrow\{\Initial(L,P_0,\Sigma)\}$
  \nllabel{ln:lifetime-init}\;
$\mathcal E\leftarrow\emptyset$\;

\While{$W\ne\emptyset$}{
  $(e,a,W)\leftarrow\Next(W)$
    \nllabel{ln:lifetime-next}\;

  $e\leftarrow\Precheck(e,a,\Sigma)$
    \nllabel{ln:lifetime-precheck}\;
  $e\leftarrow\Apply(e,a,\Sigma)$
    \nllabel{ln:lifetime-support}\;
  $e\leftarrow\Borrow(e,a,\Sigma)$
    \nllabel{ln:lifetime-borrow}\;

  \eIf{\Terminal(e)}{
    $\mathcal E\leftarrow\mathcal E\cup\{e\}$
      \nllabel{ln:lifetime-terminal}\;
  }{
    $W\leftarrow W\cup\{e\}$\;
  }
}

\Return{$\mathcal E$}
  \nllabel{ln:lifetime-return}\;
\end{algorithm}
\end{minipage}
\end{figure}

Algorithm~\ref{alg:check-lifetime} implements these two steps.
$\mathsf{InitialExecution}$ constructs the initial symbolic state from $L$,
$P_0$, and $\Sigma$ (line~\ref{ln:lifetime-init}).
$\mathsf{NextTransition}$ selects a pending path and its next action, including
one possible outcome of that action (line~\ref{ln:lifetime-next}).
For the selected action, $\mathsf{AppendPrechecks}$ records the point
requirements that must hold before the action and any exit requirement
attached to the transition (line~\ref{ln:lifetime-precheck}).
$\mathsf{ApplyAndCheckSupport}$ applies the action's ordered effects and
reevaluates active borrows whose support may have changed
(line~\ref{ln:lifetime-support}).
$\mathsf{CheckBorrowTransition}$ records borrow creation, use, transfer, or
consumption while preserving the captured object and its check history
(line~\ref{ln:lifetime-borrow}).
$\mathsf{Terminal}$ adds completed paths to $\mathcal E$; other paths return
to the worklist (line~\ref{ln:lifetime-terminal}).

The check order follows the lifetime property in Section~\ref{sec:problem}. A
point requirement must hold before its action: an acquisition cannot make its
invalid input safe. An interval requirement must hold after every effect that
can change support, not only when the borrow begins or ends. The object
captured by a borrow remains fixed even if a variable or support location
later contains another object. Once a point or interval obligation fails, a
later state change cannot repair that earlier execution prefix.
Each query uses $P_0$, the conditions established along the executed path, and
the current store $\sigma$. State-changing actions therefore affect later
queries, while $\Delta$ contributes a fact only when executed code establishes
it. Missing operands, unresolved effects, or undecided queries leave the
affected obligation incomplete.
A feasible execution prefix that fails a point, interval, or exit obligation
is retained as a \emph{violation witness}.

\noindent\textbf{Running example.}
For the bound target in Figure~\ref{fig:attribute-running}, symbolic execution
begins with slot $d$ supporting the captured payload $p$. The subsequent
$\mathsf{release}(r)$ can invoke the recovered callback. The callback replaces
the value in $d$ and releases the old payload, reducing $p$'s owner count to
zero. Because the borrow of $p$ remains active, the support check fails.
Execution then reaches $\mathsf{acquire}(p)$. Its point requirement is checked
before acquisition and also fails because $p$ is no longer live. Both failures
remain in the target path record. In the reference, $\mathsf{acquire}(p)$
executes before $\mathsf{release}(r)$ and creates another owner, so the same
callback leaves $p$ live and supported.
Lifetime checking therefore produces a safe reference path and a target path
with two failed obligations: loss of support during the borrow and an invalid
use at $\mathsf{acquire}(p)$.

\subsection{Reference Comparison}\label{sec:comparison}
\label{sec:conditional-correctness}

Reference comparison determines what the target must return to its caller and
whether each target execution has a safe reference counterpart. As
Figure~\ref{fig:technique-overview} shows, it has two steps: recover caller
obligations and match reference executions.

\noindent\textbf{Recover caller obligations.}
{\tech} executes the reference program $L_R$ under caller premises $P_0$.
For each reference path, it follows ownership changes to the exit and records
whether the returned handle is owned or borrowed. The recovered contract
$\kappa$ associates each proved output role with the input domain on which it
holds.
The target must satisfy the same role on that domain. If the reference returns
an owned handle, for example, the target must transfer ownership to its caller.
Returning a live borrow does not satisfy this requirement. If no single role
can be established for part of the input domain, that part remains unresolved.

\noindent\textbf{Match reference executions.}
{\tech} checks the reference and target under $\kappa$. For each target path,
it searches for a safe reference path with related inputs, runtime outcomes
permitted by $\Sigma$, and a matching caller-visible exit under $\rho$. Such a
path is a \emph{reference witness}. Matching compares returned objects,
aliases, ownership, borrow support, exit kind, and ordered visible effects.
A target path has \emph{reference coverage} when its witnesses cover every
input and runtime outcome admitted by that path. A failed point, interval, or
exit obligation produces a candidate violation. A target execution without a
reference witness is recorded as a coverage gap.

\begin{figure}[htp]
\centering
\ifdefined\Description
\Description{Reference comparison recovers caller obligations, checks the
reference and target under those obligations, matches safe reference
executions to target executions, and replays failed obligations before
reporting them.}
\fi
\begin{minipage}{0.7\linewidth}
\AlgorithmMinipageSetup{\footnotesize}
\begin{algorithm}[H]
\setlength{\algomargin}{0.35em}
\DontPrintSemicolon
\LinesNumbered
\caption{Recover caller obligations and match reference executions}
\label{alg:compare-protocol}

\SetKwFunction{Recover}{RecoverCallerObligations}
\SetKwFunction{Check}{CheckPrograms}
\SetKwFunction{Match}{MatchReferenceExecutions}
\SetKwFunction{Replay}{ReplayViolations}
\SetKw{Return}{return}

\KwIn{$b=(L_R,L_T,\rho,P_0,\Delta,\Lambda)$: bound comparison;
$\Sigma$: runtime specification}
\KwOut{$B$: violation reports; $Q$: coverage gaps}

$\kappa\leftarrow\Recover(L_R,P_0,\Sigma)$
  \nllabel{ln:compare-contract}\;

$(R,T)\leftarrow\Check(L_R,L_T,\kappa,P_0,\Sigma)$
  \nllabel{ln:compare-check}\;

$Q\leftarrow\Match(R,T,\rho,\Sigma)$
  \nllabel{ln:compare-match}\;

$B\leftarrow\Replay(T,b,\Sigma)$
  \nllabel{ln:compare-replay}\;

\Return{$(B,Q)$}
  \nllabel{ln:compare-return}\;
\end{algorithm}
\end{minipage}
\end{figure}

Algorithm~\ref{alg:compare-protocol} implements these two steps.
$\mathsf{RecoverCallerObligations}$ executes the reference and derives
$\kappa$ from the output role proved on each input domain
(line~\ref{ln:compare-contract}).
$\mathsf{CheckPrograms}$ checks both programs under these requirements and
produces their checked paths $R$ and $T$ (line~\ref{ln:compare-check}).
$\mathsf{MatchReferenceExecutions}$ finds safe reference witnesses for target
paths and records uncovered executions in $Q$ (line~\ref{ln:compare-match}).
$\mathsf{ReplayViolations}$ replays failed target obligations using their
recorded execution choices and returns the reproduced failures in $B$
(line~\ref{ln:compare-replay}).

The two steps check different parts of the property in
Section~\ref{sec:problem}. Recovering $\kappa$ establishes the ownership
required at the caller boundary. A borrowed result may remain live throughout
the target yet violate an interface that requires an owned result. Reference
matching checks the relational requirement: the target's caller-visible result
must correspond to one produced by a safe reference execution.
Let $t$ be a target path, $i$ an entry satisfying $P_0$, and $\eta_T$ the
runtime outcomes along $t$. Let $\mathsf{Witness}(r,t,i,\eta_T)$ mean that
reference path $r$ is safe, its runtime outcomes are permitted by $\Sigma$,
and its inputs and caller-visible exit match $t$ under $\rho$. Reference
coverage requires
\[
  \forall i,\eta_T:\quad
  P_0(i)\land\chi_t(i,\eta_T)
  \Longrightarrow
  \exists r.\ \mathsf{Witness}(r,t,i,\eta_T),
\]
where $\chi_t$ is the path condition of $t$. Different target inputs may use
different reference witnesses. A target execution without such a witness
remains uncovered.

\noindent\textbf{Running example.}
For the attribute rewrite in Figure~\ref{fig:attribute-running}, the reference
returns an owned handle to the original payload $p$. {\tech} therefore
recovers an owned-result requirement in $\kappa$. A target that returns only a
borrowed $p$ violates this requirement even if the borrow remains supported.
For the unsafe target path, lifetime checking has already found a loss of
support followed by an invalid acquisition. The corresponding reference path
acquires $p$ before releasing $r$, remains safe through the callback, and
returns the required owner. It therefore provides a reference witness for the
same input and callback outcome. Replay reproduces the target's failed
obligations before reporting them.

\noindent\textbf{Conditional correctness of completed checks.}
For a bound comparison $b=(L_R,L_T,\rho,P_0,\Delta,\Lambda)$, a check is
\emph{completed} when the programs and interfaces are constructed, all actions
and effects relevant to the obligations have usable bindings in $\Sigma$, all
admitted paths are explored, all required point, interval, and exit queries
are decided, and reference coverage is complete. Unsupported control flow,
unsummarized cycles, unresolved source relations, missing bindings, and
undecided queries make the dependent check \emph{incomplete}. A completed
check is \emph{discharged} when all its obligations hold.
For a discharged comparison, every target execution admitted by $P_0$ and
$\Sigma$ satisfies $\operatorname{Req}_{\Gamma}$ at each use,
$\operatorname{Sup}_{\Gamma}$ throughout every active borrow interval, and
$\operatorname{Exit}_{\Gamma}$ at every exit. Hence it satisfies
$\operatorname{Safe}_{\Gamma}$. Every terminating target execution also has a
safe, terminating reference witness whose caller-visible observation matches
under $\rho$. This establishes the relational property of
Section~\ref{sec:problem} for $L_R$ and $L_T$. The premises are $P_0$ and facts
established by executed code; optimizer proposals $\Delta$ are not assumed.
The result carries to corresponding concrete executions when source recovery
faithfully represents the selected actions and interfaces, and each runtime
binding faithfully realizes its declared preconditions, effects, outcomes, and
exits. Under these conditions, the symbolic states and reference witnesses
correspond to concrete runtime behavior.

\emph{Proof sketch.}
Algorithm~\ref{alg:check-lifetime} checks each point requirement before its
action. It checks borrow support at creation and after each ordered effect that
can change the support footprint. Frame conditions preserve support across
disjoint effects, and transfer preserves the active interval and prior checks.
Exit checks establish the caller's ownership requirements on every modeled
exit. Discharge over all admitted paths therefore establishes
$\operatorname{Safe}_{\Gamma}$.
Algorithm~\ref{alg:compare-protocol} establishes reference coverage for every
admitted input and target outcome, yielding a safe matching reference witness
for each terminating target execution. These conditions establish the
relational preservation property for the completed comparison.

\section{Implementation}\label{sec:implementation}

We implemented {\tech} with a C++17 source frontend, a Python symbolic checker,
and Z3 4.13.0~\cite{z3-solver}. The frontend recovers rewrites and their source
context from a CPython source tree. The Python checker binds recovered actions
to runtime contracts, executes the resulting programs, checks lifetime
obligations, and compares reference and target executions.

\vspace{3pt}
\noindent\textbf{Source Recovery and Binding.}
The frontend analyzes the C implementation of runtime operations together with
the Python code that constructs optimizer rewrites. Clang-based
analysis~\cite{clang-libclang} connects definitions, uses, calls, and
pointer-valued expressions. Andersen-style points-to analysis supplies
conservative pointer relations~\cite{andersen-pta,grove-callgraph}.
For each optimizer path, the frontend reconstructs operation-buffer updates in
execution order to recover the selected reference and target programs. It also
retains the path condition, operands, and source code that supplies inputs and
receives exits. Ownership-relevant fields remain separate abstract locations,
and unresolved actions remain explicit.
The frontend exports these results to the Python checker. The lowerer binds
each recovered action to a runtime contract that specifies its preconditions,
ordered state changes, and possible outcomes. Actions with missing operands or
unsupported runtime effects remain opaque.

\vspace{3pt}
\noindent\textbf{Lifetime Checking.}\label{sec:execution-details}
The symbolic kernel explores each bound program with a worklist of path states.
A state contains the path condition, symbolic object and storage state, active
borrows, completed checks, and exit state. State-changing actions create new
symbolic versions, so each obligation is checked against the state at that
program point.
A borrow record retains the captured object, its supporting locations, and the
support predicate's read footprint. The kernel checks point obligations before
their actions, reevaluates active support after effects that may change it, and
checks ownership and transfer requirements at exits. Borrow transfers preserve
the captured object and prior check history.
Z3 decides branch feasibility, obligation predicates, alias relations, and
footprint overlap. Optimizer proposals remain separate from facts established
by executed code. Unsupported relations and undecided queries leave the
affected check incomplete.

\vspace{3pt}
\noindent\textbf{Reference Comparison.}
The same kernel executes reference paths to recover their output roles and
input domains. These results form the caller contract $\kappa$. The checker
then executes both reference and target programs under $\kappa$.
For each target path, {\tech} searches for a safe reference path with related
inputs, runtime outcomes allowed by the contracts, and a matching
caller-visible exit under $\rho$. Matching compares returned objects, aliases,
ownership, borrow support, exit kind, and ordered visible effects. Z3 checks
whether the accepted reference witnesses cover every admitted target input and
outcome choice. Executions without complete reference coverage are recorded as
coverage gaps.

\vspace{3pt}
\noindent\textbf{Reports and Replay.}\label{sec:replay-details}
For each completed failure, {\tech} records the affected source operation,
failed obligation, path condition, runtime outcomes, and ordered effects.
Before exporting a report, replay reconstructs the recovered program, bindings,
symbolic path, and obligation query from the recorded inputs. A violation is
reported when replay reproduces the failure.
Missing contracts, opaque actions, unresolved paths, incomplete reference
coverage, and undecided queries remain internal completeness records. Replay
checks the consistency of a reported witness. Source review and native
execution determine whether the modeled behavior occurs in the concrete
runtime.

\section{Evaluation}\label{sec:evaluation}

We evaluate {\tech} along four questions:
\begin{description}[style=unboxed,leftmargin=0pt,nosep]
\item[\textbf{RQ1:}] How accurately and completely does {\tech} detect lifetime violations?
\item[\textbf{RQ2:}] What are its analysis costs?
\item[\textbf{RQ3:}] Which design components contribute to detected violations?
\item[\textbf{RQ4:}] What defects does {\tech} find, and what causes and runtime effects do they have?
\end{description}

\subsection{Experimental Setup}\label{sec:eval-setup}

\noindent\textbf{Subjects and benchmarks.}
We evaluate controlled accuracy, mechanism coverage, production behavior, and
frozen-model transfer.
\PyJITBench{} is our controlled benchmark suite for lifetime and ownership
properties in JIT rewrites. It contains portable C cases, JIT-shaped rewrites,
protocol controls, and source-level comparisons covering ownership and output
contracts, guards, callbacks and finalization, helper effects, stack handoffs,
side exits, and runtime bindings.
The primary controlled benchmark contains 58 portable C inputs: 12 modified
C-API cases and 46 encodings of 23 historical mechanisms. Their registered
defect/repair labels give 29 defective and 29 safe inputs. Case selection and
labels were established independently of {\tech} output. The historical
encodings retain source revisions and repair rationales, while the C-API cases
specify their ownership conditions. All three tools analyze the same C inputs.
The benchmark was later used during implementation. Additional authored-C,
JIT-shaped, protocol, and source-semantic controls exercise individual
mechanisms.
For production evaluation, we analyze all rewrites extracted from
CPython~3.14.8, 3.15.0, and a 3.16 development snapshot archived on
18 October 2026~\cite{python-3148-release,python-3150-release,cpython-316-snapshot}.
Across the three snapshots, the frontend extracts 3,332 rewrite alternatives
before feasibility filtering. We use this population to measure report
precision and semantic coverage. Revision, transfer, runtime-binding, and
scaling studies use separate inputs described with their experiments. The
frozen revision studies apply a fixed checker and fixed contracts to
separately selected cases.

\noindent\textbf{Baselines and configurations.}
We compare {\tech} with PyRefcon~\cite{pyrefcon} and
CpyChecker~\cite{cpychecker-docs}. All three tools analyze the same portable C
cases in the controlled comparison; {\tech} additionally uses the reference
computation and runtime contracts. On production CPython, PyRefcon and
CpyChecker analyze \path{Python/optimizer_analysis.c}, while {\tech} analyzes
the recovered rewrites.
PyRefcon runs with Clang~13 after removing two obsolete code-constructor models
for header compatibility. CpyChecker runs with GCC~6.5.0 and Python~3.6;
per-function exceptions are counted as tool errors.
For component comparisons, we use two {\tech} diagnostic configurations. The
\emph{full} configuration enables all source checks and reference--target
comparisons. The \emph{local + feasible effects} configuration uses the same
recovered paths, premises, helper outcomes, and ordered effects, but omits
warning cancellation and paired checks of outputs, aliases, guard domains, and
exit kinds. Each experiment states the configuration it uses.

\noindent\textbf{Outcomes and metrics.}
A \emph{report} is a replay-validated failure of a point, interval, or exit
obligation. Internal checks are classified as discharged (D), violated (V), or
incomplete (U). D denotes a completed comparison whose modeled obligations and
required reference coverage hold. V contains a replay-validated failure and
produces a report. U denotes a comparison with an unresolved required contract,
path, solver result, or reference witness. Construction failures are recorded
separately when no comparison is formed.
Precision is the fraction of reports confirmed by manual review; recall is the
fraction of labeled defects detected. The controlled benchmarks provide labels
for both measures. For production reports, two reviewers independently assess
all 286 reports from the broader configuration using each witness and its
source context. The final review classifies 92 reports as true and 194 as
false. The default receiving-context rule retains 194 of these reports: 92
true and 102 false. Production precision is computed on this retained set.
Production non-reports are not exhaustively labeled, so production recall is
not measured.

Receiving-context refinement was developed while analyzing the production
population. We therefore report precision for the resulting configuration on
this population. The frozen studies use separately selected cases with the
checker and contracts fixed before prediction.
We also measure semantic coverage and analysis cost. Semantic coverage records
whether an extracted rewrite has the bindings, recovered output roles, and
reference witnesses needed to complete its checks. Analysis cost includes
wall-clock time, memory use, and solver activity where reported.
A report refers to one analyzed input. A \emph{root cause} groups reports that
arise from the same source defect. Static confirmation requires a reachable
witness that establishes the assigned cause. Native reproduction additionally
requires a passing control, activation of the target code, and a
cause-specific failure oracle. Upstream confirmation requires an explicit
response from a runtime developer.

\noindent\textbf{Environment.}
Experiments run on Linux~6.8.0 with an Intel Core i9-13900F, 32 logical CPUs,
and 62.5~GiB of memory. The analysis toolchain uses GCC~9.5.0,
LLVM/libclang~14.0.0, Python~3.10.9, and Z3~4.13.0. Native
descriptor/finalizer harnesses link CPython~3.10.9, and the sequential
weak-reference experiments use GLib~2.72.4. Population analyses impose no
fixed limit on path count or exploration depth. Each native study states
whether JIT code is active.

\subsection{Effectiveness and Semantic Coverage (RQ1)}
\label{sec:eval-effectiveness}

The shared portable C benchmark measures controlled accuracy across three
tools. Additional development controls test specific mechanisms. Production
CPython measures report precision and semantic coverage on real optimizer
source. Frozen transfer studies test the fixed model on separately selected
revisions.

\subsubsection{Independently Constructed Controlled Benchmark}

The 58-input benchmark contains six defective and six safe C-API cases, 23
defective historical encodings, and 23 repaired identity controls. An identity
control compares a repaired encoding with itself; the six safe C-API cases
compare different implementations that satisfy the same contract. Case
selection and labels were fixed independently of {\tech} output, and the
benchmark was later used during development. Each tool receives the same
candidate C source; {\tech} additionally receives the reference and runtime
contracts.

On the 58 portable inputs shared by all three tools, {\tech} detects all 29
labeled defects and reports none of the 29 safe inputs. PyRefcon detects 20
defects and CpyChecker 18; neither reports a safe input. {\tech} therefore
achieves 100\% precision and recall on this population, compared with 68.97\%
recall for PyRefcon and 62.07\% for CpyChecker
(Table~\ref{tab:controlled-accuracy}).

\begin{table*}[!htbp]
\centering
\footnotesize
\caption{Shared controlled accuracy and additional mechanism controls.
``Detected'' uses defective inputs as the denominator; ``No report'' uses safe
inputs. Errors are reported separately. PyRefcon and CpyChecker do not support
the JIT-shaped interface.}
\label{tab:controlled-accuracy}
\setlength{\tabcolsep}{2.5pt}
\begin{tabular*}{\linewidth}{@{\extracolsep{\fill}}lrrrrrrr@{}}
\toprule
\makecell{Inputs} &
\makecell{Tool} &
\makecell{Detected} &
\makecell{No report} &
\makecell{FP} &
\makecell{Errors} &
\makecell{Precision} &
\makecell{Recall} \\
\midrule
\multicolumn{8}{l}{\textit{Primary controlled accuracy comparison}} \\
Shared C &
{\tech} &
29/29 & 29/29 & 0 & 0 & 100.00\% & 100.00\% \\
Shared C &
PyRefcon &
20/29 & 29/29 & 0 & 0 & 100.00\% & 68.97\% \\
Shared C &
CpyChecker &
18/29 & 29/29 & 0 & 0 & 100.00\% & 62.07\% \\
\midrule
\multicolumn{8}{l}{\textit{Additional development controls}} \\
Authored C &
{\tech} &
16/16 & 16/16 & 0 & 0 & 100.00\% & 100.00\% \\
Authored C &
PyRefcon &
8/16 & 16/16 & 0 & 0 & 100.00\% & 50.00\% \\
Authored C &
CpyChecker &
11/16 & 7/16 & 7 & 2 & 61.11\% & 68.75\% \\
\midrule
JIT-shaped &
{\tech} &
6/6 & 6/6 & 0 & 0 & 100.00\% & 100.00\% \\
\bottomrule
\end{tabular*}
\end{table*}

\rqfinding{On the 58 shared C inputs, {\tech} detects all 29 defects and reports
none of 29 safe inputs (100\% precision and recall). PyRefcon detects 20
defects and CpyChecker 18; neither reports a safe input.}

\subsubsection{Additional Mechanism Controls}

The authored-C benchmarks cover additional source patterns while retaining
labeled defective and changed-safe pairs. {\tech} detects all 16 defects and
reports none of the 16 safe inputs. PyRefcon detects 8 defects and reports none
of the safe inputs. CpyChecker detects 11 defects, reports seven safe inputs,
and encounters two tool errors. The 12 JIT-shaped inputs exercise the rewrite
interface unavailable to the two baselines; {\tech} detects all six defects
and reports none of the six safe rewrites.

Additional controls test the JIT-specific mechanisms used by the analysis.
The protocol checker classifies 20 controls as violated, 20 as discharged, and
six intentionally incomplete cases as incomplete, matching all registered
expectations. On 18 source-level comparisons covering C-API ownership, runtime
producers, descriptors, and final-release callbacks, it produces nine reports
and nine non-reports; five non-reports complete and four remain incomplete.
Every reported witness replays.
These controls exercise the three design stages under known semantics. Source
recovery preserves rewrite order and source context (C1); lifetime checking
checks ownership and support at the required program points (C2); reference
comparison checks returned ownership and caller-visible agreement (C3). These
development controls therefore measure mechanism coverage separately from the
58-input accuracy benchmark.

\subsubsection{Production CPython}

Across CPython~3.14.8, 3.15.0, and 3.16-dev, {\tech} extracts 3,332 rewrite
alternatives and filters 2,531 as infeasible under recovered receiving
contexts. Among the remaining 801 alternatives, it reports 194 violations.
Manual review classifies 92 reports as true and 102 as false, giving
$92/194=47.42\%$ precision.

\begin{table*}[!htbp]
\centering
\footnotesize
\caption{Production results by CPython version. {\tech} reports recovered
rewrites; PyRefcon reports translation-unit warnings; CpyChecker reports
function-level warnings. Production non-reports are not exhaustively labeled,
so the runs have no common recall denominator.}
\label{tab:production-comparison}
\setlength{\tabcolsep}{3pt}
\begin{tabular*}{\linewidth}{@{\extracolsep{\fill}}llrrrr@{}}
\toprule
\textbf{Version} &
\textbf{Tool} &
\textbf{Reports} &
\textbf{True} &
\textbf{False} &
\textbf{Errors} \\
\midrule
3.14.8 &
{\tech}   & 2   & 0  & 2  & 0 \\
3.14.8 &
PyRefcon   & 2   & 0  & 2  & 0 \\
3.14.8 &
CpyChecker & 8   & 0  & 8  & 1 \\
\midrule
3.15.0 &
{\tech}   & 96 & 46 & 50 & 0 \\
3.15.0 &
PyRefcon   & 5   & 0  & 5   & 0 \\
3.15.0 &
CpyChecker & 9   & 0  & 9   & 0 \\
\midrule
3.16-dev &
{\tech}   & 96 & 46 & 50 & 0 \\
3.16-dev &
PyRefcon   & 5   & 0  & 5   & 0 \\
3.16-dev &
CpyChecker & 9   & 0  & 9   & 0 \\
\midrule
\multicolumn{2}{l}{\textbf{{\tech} total}} &
\textbf{194} & \textbf{92} & \textbf{102} & \textbf{0} \\
\multicolumn{2}{l}{\textbf{PyRefcon total}} &
\textbf{12} & \textbf{0} & \textbf{12} & \textbf{0} \\
\multicolumn{2}{l}{\textbf{CpyChecker total}} &
\textbf{26} & \textbf{0} & \textbf{26} & \textbf{1} \\
\bottomrule
\end{tabular*}
\end{table*}

We also run PyRefcon and CpyChecker on the corresponding optimizer sources.
Table~\ref{tab:production-comparison} reports the three tools by CPython
version. {\tech} analyzes recovered reference--target rewrites, PyRefcon
analyzes the optimizer translation unit, and CpyChecker analyzes functions in
that unit. The production runs therefore compare behavior on the same source
versions across different analysis units.
{\tech} produces no true production report on CPython~3.14.8 and 46 on each
of 3.15.0 and 3.16-dev. PyRefcon emits 2, 5, and 5 warnings across the same
versions, while CpyChecker emits 8, 9, and 9. Manual source review classifies
all baseline warnings as false. The controlled benchmark provides the direct
three-tool detection comparison; these production runs show how each tool
behaves on real optimizer source.

\begin{figure*}[tbp]
\centering
\includegraphics[width=\linewidth]{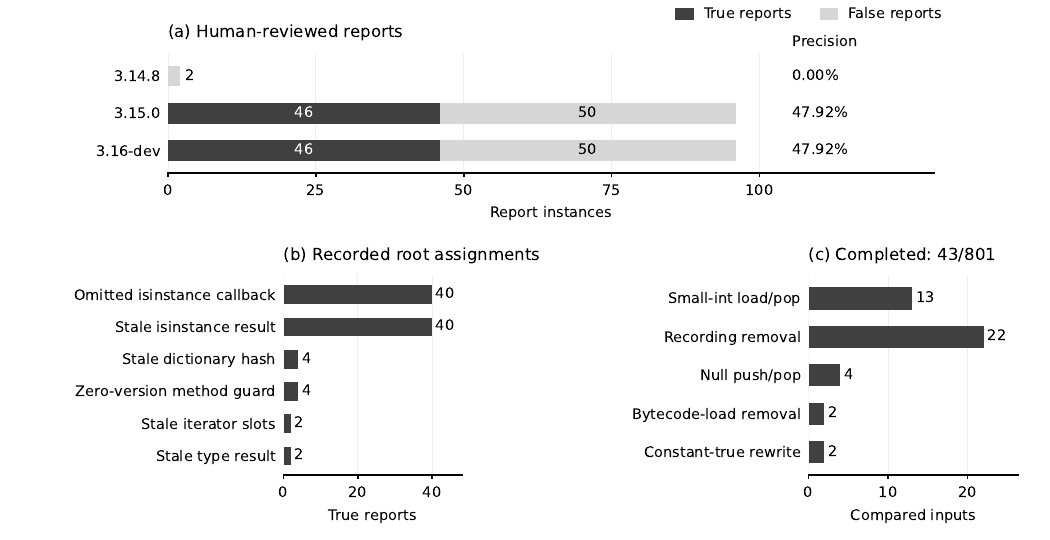}
\caption{{\tech} production results: (a) true and false reports by version;
(b) root assignments of true reports; and (c) the 43 completed non-reporting
checks among 801 feasible alternatives. Counts include alternatives repeated
across versions.}
\Description{Grayscale bars compare reviewed {\tech} reports, their root
assignments, and completed non-reporting checks by rewrite family.}
\label{fig:production-results}
\end{figure*}

The production run also measures semantic coverage. Among 607 feasible
alternatives that produce no report, 43 complete and 564 remain incomplete
(Figure~\ref{fig:production-results}). The 801 feasible alternatives therefore
consist of 194 reports, 43 completed non-reports, and 564 incomplete
non-reports.
Most incomplete cases arise from missing source or runtime bindings, unresolved
control flow and stack handoffs, missing helper effects, or incomplete
reference witnesses. These gaps map directly to the three design stages. C1
must recover and bind the actions required by the rewrite; C2 needs their
effects to complete lifetime checks; and C3 needs enough reference behavior
to establish coverage. None of the recorded reference-coverage queries ends
with an undecided solver result. Missing modeled semantics therefore dominate
production incompleteness.

The baseline warnings also expose gaps in ownership modeling. PyRefcon reports
cases involving \path{Py_DECREF}, immortal singleton objects, and ownership
transfer in \path{sym_new_const_steal}. CpyChecker similarly misclassifies
several borrowed frame, type, tuple, lookup, and symbol pointers or misses the
same ownership transfer. {\tech} uses recovered rewrite context and explicit
runtime contracts. Its 47.42\% production precision also shows that modeling
precision remains important for production analysis.

\rqfinding{Across 801 feasible production rewrites, {\tech} reports 194
violations, of which 92 are true (47.42\% precision). PyRefcon emits 12
warnings and CpyChecker 26 on the corresponding optimizer sources, all
classified as false by manual review. Among {\tech}'s 607 non-reporting
rewrites, 43 complete and 564 remain incomplete, primarily because required
source, runtime, or reference semantics are missing.}

\subsubsection{Frozen Transfer and Model Coverage}

We freeze the {\tech} checker and runtime contracts before evaluating three
separately selected revision populations. These studies test how the fixed
model transfers beyond the controlled benchmark. The populations contain
three cases with native defect labels, so we report their outcomes and failure
modes at the case level.
The changed-runtime-function study contains 10 pairs, including one defect
labeled by native execution before prediction. The beta study contains 13
pairs with two native-tested regressions. {\tech} misses all three labeled
defects. The historical window contains 105 pairs, including 96 preservation
labels; their identity controls retain the same reports.

\begin{table}[!htbp]
\centering
\footnotesize
\caption{Frozen transfer and model coverage. All three native-labeled defects
are missed. Preservation labels state equivalence of the selected versions.}
\label{tab:transfer-main}
\begin{tabularx}{\linewidth}{@{}p{0.22\linewidth}XX@{}}
\toprule
\textbf{Population} &
\textbf{Pre-prediction evidence} &
\textbf{Observed result} \\
\midrule
Changed runtime functions &
10 pairs; one native defect label &
{\tech} misses the defect \\
Historical window &
105 pairs; 96 preservation labels &
Identity controls retain the same reports \\
Beta window &
13 pairs; two native regressions &
{\tech} misses both regressions \\
\bottomrule
\end{tabularx}
\end{table}

The three misses identify gaps in the frozen model. The dictionary case
reaches operations and a \path{PyDict_Check} condition absent from the model.
The beta cases require caller or callback semantics that the model does not
provide. These missing semantics prevent source recovery or reference
comparison from constructing complete lifetime and reference witnesses.
The controlled benchmark measures detection under explicit benchmark
semantics. The frozen study shows which source and runtime semantics are
needed to extend the fixed model to the selected revisions.

\rqfinding{With the checker and runtime contracts frozen, {\tech} misses all
three native-labeled regressions in the transfer populations. The misses
require additional source, caller, or callback semantics, identifying model
coverage as the main transfer limitation.}

\subsection{Efficiency (RQ2)}\label{sec:eval-efficiency}

We measure end-to-end production cost and then vary program structure to
identify the main sources of analysis cost.

\subsubsection{Production Cost}

Full production analysis takes 35.56--42.31~s per CPython snapshot, with peak
resident memory of 613.6--830.9~MiB
(Table~\ref{tab:analysis-cost-overview}). These runs use canonical receiving
contexts and analyze the RQ1 production population. The weakened-context runs
use the broader population from the source-recovery ablation. The unfiltered
runs retain every extracted alternative and are substantially more expensive
on CPython~3.15.0 and 3.16-dev.

\begin{table*}[!htbp]
\centering
\footnotesize
\caption{Analysis cost. Protocol time is per CPython snapshot. Shared-C time
is the total over 58 inputs and includes process launch and adaptation. RSS is
the largest single-process resident set.}
\label{tab:analysis-cost-overview}
\setlength{\tabcolsep}{2.5pt}
\begin{tabular*}{\linewidth}{@{\extracolsep{\fill}}lrrrr@{}}
\toprule
\makecell{Population} &
\makecell{Version or tool} &
\makecell{Measurement} &
\makecell{Wall s} &
\makecell{Max RSS (MiB)} \\
\midrule
Full                & 3.14.8   & 1 run / 16 workers & 35.56          & 613.6 \\
Full                & 3.15.0   & 1 run / 16 workers & 41.32          & 830.9 \\
Full                & 3.16-dev & 1 run / 16 workers & 42.31          & 809.1 \\
\midrule
Weakened context    & 3.14.8   & 1 run / 16 workers & 34.86          & 615.0 \\
Weakened context    & 3.15.0   & 1 run / 16 workers & 51.06          & 827.9 \\
Weakened context    & 3.16-dev & 1 run / 16 workers & 54.32          & 811.8 \\
\midrule
Protocol, unfiltered & 3.14.8   & 2 runs / 32 workers & 47.28--52.62   & 1032.0 \\
Protocol, unfiltered & 3.15.0   & 2 runs / 32 workers & 576.32--583.12 & 1478.5 \\
Protocol, unfiltered & 3.16-dev & 2 runs / 32 workers & 580.54--580.83 & 1453.2 \\
\midrule
58 shared C inputs   & {\tech}   & 1 run / input & 1.05 & 73.2 \\
58 shared C inputs   & PyRefcon   & 1 run / input & 2.40 & 60.9 \\
58 shared C inputs   & CpyChecker & 1 run / input & 8.46 & 152.9 \\
\bottomrule
\end{tabular*}
\end{table*}

The full and weakened-context runs use the same checker and 16 workers, with
one timed run per snapshot. Canonical context recovery removes alternatives
before symbolic execution. Relative to weakened context, it reduces elapsed
time on 3.15.0 and 3.16-dev and adds 0.70~s on 3.14.8. The unfiltered runs use
32 workers and retain all extracted alternatives.
On the 58 portable inputs shared by all three tools, {\tech} takes 1.05~s in
total, compared with 2.40~s for PyRefcon and 8.46~s for CpyChecker. These
inputs provide a direct cost comparison because all three tools receive the
same candidate programs. For the source-level {\tech} controls, construction
takes 0.030--0.815~s, lowering and certificate generation take
0.159--2.646~s, and replay takes 0.140--4.425~s. The largest certificate is
2.52~MiB.

\rqfinding{Full production analysis completes in 35.56--42.31~s with
613.6--830.9~MiB peak RSS per CPython snapshot. On the 58 inputs shared by all
three tools, {\tech} takes 1.05~s, compared with 2.40~s for PyRefcon and
8.46~s for CpyChecker.}

\begin{figure*}[!tb]
\centering
\includegraphics[width=\textwidth]{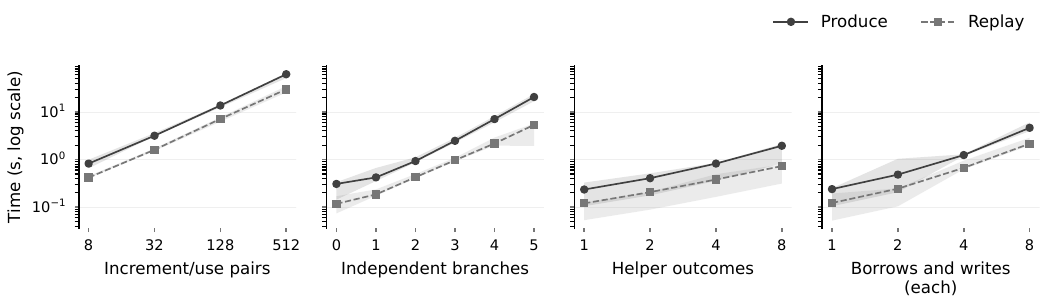}
\caption{Protocol-checker cost as input structure grows. Points show medians
of five measured runs; bands show observed ranges. Time uses a logarithmic
scale. }
\Description{Four grayscale plots compare analysis and replay cost as
straight-line actions, branches, helper outcomes, and active borrows increase.}
\label{fig:protocol-costs}
\end{figure*}

\subsubsection{Scaling Behavior}

We vary four structures that directly affect the analysis: trace length,
independent branches, helper outcomes, and simultaneously active borrows.
Eighteen generated identity pairs contain 8--512 increment/use pairs, up to
five independent Boolean branches (32 paths), one--eight helper outcomes, or
one--eight active borrows with intervening writes. Each input is analyzed
without a path, solver, state, or time cutoff. All 108 measured runs complete
without a violation, and every certificate replays.

Figure~\ref{fig:protocol-costs} shows how analysis and replay costs change with
each input structure. At 512 increment/use pairs, median analysis and replay
take \EvalDesignActionProduce{}~s and \EvalDesignActionReplay{}~s and issue
4,099 and 2,050 solver queries, respectively. At 32 branch paths, they take
\EvalDesignBranchProduce{}~s and \EvalDesignBranchReplay{}~s, with 1,566 and
352 queries. Eight helper outcomes take \EvalDesignOutcomeProduce{}~s and
\EvalDesignOutcomeReplay{}~s; eight active borrows take
\EvalDesignBorrowProduce{}~s and \EvalDesignBorrowReplay{}~s. The largest
certificate is 268,654 bytes, and the largest measured process peak is
\EvalDesignPeakMemory{}~MiB.
These results follow the structure of C2 and C3. Longer traces increase the
number of program states and obligation queries. Independent branches create
more execution paths and enlarge the reference-witness search in C3. Replay is
cheaper in branch-heavy cases because it follows recorded paths and witnesses.
Helper outcomes add runtime alternatives, while active borrows add support
checks; neither produces the same path multiplication as independent branches.

\rqfinding{Analysis cost grows most with long traces and independent branches.
Long traces increase solver work, while branches increase both path
exploration and reference-witness search; replay follows recorded witnesses
and avoids that search.}

\subsection{Design Contribution Analysis (RQ3)}
\label{sec:eval-ablations}

RQ3 measures how the mechanisms introduced for C1--C3 affect report detection
and semantic coverage. Each experiment holds its inputs and contracts fixed
while removing, weakening, or replacing one mechanism.
Tables~\ref{tab:context-sensitivity}, \ref{tab:lifetime-ablation}, and
\ref{tab:reference-ablation} summarize the report-level comparisons; the
following sections interpret them by design stage.

\subsubsection{Source Recovery}

Full source recovery reconstructs all 36 expected source relations and all four
ordered-buffer controls. Restricting recovery to direct calls misses three
targets and two data-flow relations, while field-insensitive recovery adds one
false target. Both call-target precision and indirect relation recovery
therefore affect construction of the intended rewrite.
The production context experiment shows a second effect of source recovery.
On the same 3,332 extracted alternatives, the full configuration uses canonical
receiving contexts. Weakening this recovery retains the mandatory
\path{_CALL_ISINSTANCE} entry contract but omits canonical-context filtering
for the other operation families. The feasible population grows from 801 to
912, adding 92 reviewed false reports and 19 incomplete non-reports. The 92
true reports and 43 completed non-reports are unchanged, while precision falls
from 47.42\% to 32.17\%.

\begin{table}[!htbp]
\centering
\footnotesize
\caption{Effect of receiving-context recovery on the production population.}
\label{tab:context-sensitivity}
\begin{tabular*}{\linewidth}{@{\extracolsep{\fill}}lrrrrr@{}}
\toprule
\textbf{Context} &
\textbf{Feasible} &
\textbf{Reports} &
\textbf{True} &
\textbf{False} &
\textbf{Precision} \\
\midrule
Full             & 801 & 194 & 92 & 102 & 47.42\% \\
Weakened context & 912 & 286 & 92 & 194 & 32.17\% \\
\bottomrule
\end{tabular*}
\end{table}

These measurements isolate the contribution of C1. Source recovery must
preserve operation order, data-flow relations, and the context in which the
rewrite is reachable. On the production population, canonical context recovery
removes infeasible candidates and improves precision while leaving the true
reports unchanged.

\rqfinding{Source recovery reconstructs all 36 expected relations and four
ordered-buffer controls. Restricting recovery loses three targets and two
relations, while canonical receiving context removes 92 reviewed false
production reports without removing a true report.}

\subsubsection{Lifetime Checking}

Temporal lifetime reasoning has the largest effect on defect detection.
Endpoint-only borrow checking reduces ten reports to six, missing four
violations in which support disappears and is restored before the next
endpoint. Freezing the original support locations loses one additional report,
and omitting operand equality reduces seven reports to five.
Point checks contribute independently. Omitting action preconditions reduces
three reports to two, while retaining only the latest point check reduces six
reports to four. Assuming optimizer proposals as established facts removes
both reports in its controls.
Path conditions affect precision. Omitting branch conditions increases the
report count from seven to twelve, adding five false reports from infeasible
executions.

\begin{table}[!htbp]
\centering
\footnotesize
\caption{Selected lifetime-checking ablations.}
\label{tab:lifetime-ablation}
\begin{tabularx}{\linewidth}{@{}Xrrr@{}}
\toprule
\textbf{Change} & \textbf{Full} & \textbf{Variant} & \textbf{$\Delta$} \\
\midrule
Assume optimizer proposals    & 2  & 0  & $-2$ \\
Omit preconditions            & 3  & 2  & $-1$ \\
Check borrow endpoints only   & 10 & 6  & $-4$ \\
Freeze support locations      & 1  & 0  & $-1$ \\
Omit operand equality         & 7  & 5  & $-2$ \\
Keep latest point check only  & 6  & 4  & $-2$ \\
Omit branch conditions        & 7  & 12 & $+5$ \\
\bottomrule
\end{tabularx}
\end{table}

These results separate the contributions of C2. Point checks detect invalid
uses at individual operations, while interval checks detect temporary support
loss between them. Operand equality and support-location tracking preserve the
identity of the borrowed object and its support. Path conditions exclude
infeasible executions, and optimizer facts contribute only after executed code
establishes them.

\rqfinding{Endpoint-only checking misses four violations, assuming optimizer
proposals loses two reports, and omitting branch conditions adds five false
reports. Continuous support, executed facts, and path feasibility each affect
the checker outcome.}

\subsubsection{Reference Comparison}

Caller-visible obligations expose violations that target-local checking misses.
Withholding the recovered output role changes one report to zero. Omitting exit
requirements reduces three reports to one, losing violations for a missing
acquisition and a nullable borrowed result. Restarting imported borrow history
reduces four reports to three by losing a pre-return support gap.

\begin{table}[!htbp]
\centering
\footnotesize
\caption{Selected reference-comparison ablations.}
\label{tab:reference-ablation}
\begin{tabularx}{\linewidth}{@{}Xrrr@{}}
\toprule
\textbf{Change} & \textbf{Full} & \textbf{Variant} & \textbf{$\Delta$} \\
\midrule
Withhold recovered output role  & 1 & 0 & $-1$ \\
Omit exit requirements           & 3 & 1 & $-2$ \\
Restart imported borrow history  & 4 & 3 & $-1$ \\
Check target locally only        & 3 & 3 & 0 \\
Compare output values only       & 0 & 0 & 0 \\
\bottomrule
\end{tabularx}
\end{table}

Full reference matching changes semantic coverage on its selected controls.
In the source-derived finalizer control, all 122 local target obligations pass,
yet native execution observes a different event order in 32 of 64 schedules.
Reference agreement therefore remains incomplete.
These measurements separate two contributions of C3. Recovering caller output
and exit obligations directly exposes violations. Matching complete reference
executions determines whether a locally valid target has a corresponding safe
reference execution and therefore affects semantic coverage independently of
the report count.

\rqfinding{Recovered output roles expose one additional violation, exit
requirements expose two, and imported borrow history exposes one. Full
reference matching adds no reports on its selected controls but identifies
cases where reference agreement remains incomplete.}
\subsection{Bug Types, Root Causes, and Impact (RQ4)}
\label{sec:eval-bugs}

RQ4 characterizes the failure mechanisms in the bug catalog, their connection
to the {\tech} design, and their observed runtime effects. The catalog contains
17 distinct root causes: 15 new project bugs, each with a reproducible proof
of concept, and two previously public defects.

\subsubsection{Failure Classes and Impact}

The 17 roots fall into five classes
(Table~\ref{tab:bug-causes-impact}): eight specialization-dependency failures,
three ownership/lifetime failures, and two each involving dynamic effects,
guard preconditions, and state protocols. The 92 true production reports from
RQ1 map to six roots; the remaining roots come from the diagnostic comparator
or separate source checks.

\begin{table*}[!htbp]
\centering
\footnotesize
\caption{Root-cause classes and their connection to the {\tech} design.
Counts describe distinct roots, not report instances. Impacts are those
recorded for the catalog; project-root native outcomes are summarized
separately in Table~\ref{tab:bug-evidence-ledger}.}
\label{tab:bug-causes-impact}
\setlength{\tabcolsep}{3pt}
\begin{tabularx}{\linewidth}{@{}p{0.18\linewidth}XXX@{}}
\toprule
\textbf{Class (roots)} &
\textbf{Root cause} &
\textbf{Design connection} &
\textbf{Recorded impact} \\
\midrule
Specialization dependency (8) &
Cached values lack a dependency on the current type, slot, or lookup state &
C1 binds cached values to their source objects; C2 checks that the required
type, slot, or lookup state remains valid &
Incorrect lookup and duplicate dictionary entries after a key's class changes \\

Ownership/lifetime (3) &
Release precedes acquisition, or required cleanup is omitted &
C1 preserves operation order and ownership sources; C2 checks point,
continuous-support, and exit obligations &
Stale-storage reads, retained payloads, and retained iterator types \\

Dynamic-effect omission (2) &
Optimization removes operations that may invoke callbacks or finalization &
C1 binds helper effects; C2 tracks their lifetime effects; C3 compares
caller-visible outcomes when modeled &
Changed \path{isinstance} results and omitted finalization \\

Guard precondition (2) &
A zero type-version tag reaches a guard requiring a nonzero tag &
C1 binds the emitted guard operands; C2 checks its precondition at execution &
Debug-build assertion during executor construction \\

State protocol (2) &
Exit state is stale or disagrees with the inputs consumed by the rewrite &
C1 recovers receiving-stack state; C2 checks the state required at the exit &
Extra executor creation and debug stack-depth abort \\
\bottomrule
\end{tabularx}
\end{table*}

The five classes exercise different parts of the design. Specialization
failures require {\tech} to connect cached values to the state that justifies
them. Ownership failures depend on operation order and continuous support.
Dynamic-effect failures require runtime effects not explicit in optimizer
source. Guard and state-protocol failures arise at entry or exit boundaries.
The catalog therefore spans point, interval, and exit obligations together
with the source dependencies needed to construct them.
The public defect \path{PYTHON-158804} illustrates an exit obligation: a
specialized operation consumes inputs that its async-generator side exit still
expects, causing a recorded debug-build stack-depth
abort~\cite{cpython-issue-158804}. C1 recovers the receiving-stack layout, and
C2 checks it against the consumed operands. The following two cases illustrate
specialization dependencies and continuous borrow support.

\subsubsection{Case Study 1: A Stale Hash in Dictionary Operations}
\label{sec:case-stale-hash}

In the stale-hash case, the optimizer embeds a key's hash in specialized
dictionary lookup and store operations. The test first warms these operations,
then changes the key's \path{__class__} to a compatible class with a different
\path{__hash__} implementation. It then creates a fresh dictionary containing
the same key under its new hash. The existing executor remains valid and uses
the old embedded hash on this new dictionary. The failure therefore does not
depend on changing the hash of a key already stored in that dictionary.

Figure~\ref{fig:case-stale-hash} shows the relevant source excerpts. The
optimizer computes the hash under the identity-hash condition and embeds it in
the target operation. It registers a watcher on the original type, but the
watcher does not establish that the key retains that type after
\path{__class__} reassignment. The cached hash can therefore remain in use
after the key's hash slot changes.

\begin{figure}[!htbp]
\centering
\begin{minipage}[t]{0.75\linewidth}
\vspace{0pt}
\begin{lstlisting}[style=casecode]
// Optimizer: identity-hash branch
if (Py_TYPE(key)->tp_hash == PyBaseObject_Type.tp_hash) {
    h = PyObject_Hash(key);
    ADD_OP(known_hash_op, 0, h);
    watch_type(Py_TYPE(key), deps);
}

// Target lookup
res = _PyDict_SubscriptKnownHash(dict, key, h);
\end{lstlisting}
\end{minipage}
\caption{Code excerpts for the stale dictionary hash. Excerpts from the
CPython~3.16-dev optimizer helper and lookup operation are shortened and
operands renamed; \texttt{known\_hash\_op} selects the specialized operation.
The existing type watcher is shown. The missing protection connects the
key's current type to the type used to compute the embedded hash.}
\Description{Code excerpts show hash computation, embedding, type-watcher
registration, and use by a known-hash dictionary lookup.}
\label{fig:case-stale-hash}
\end{figure}

C1 follows \path{PyObject_Hash(key)} into the emitted operand and recovers the
identity-hash selection condition. C2 checks for a guard on the same key's type
and a dependency on that type. The recorded
\path{embedded_value_stability} diagnostic identifies the missing protection
at this use. The key remains live; the failed point obligation concerns its
specialization state. Four manually reviewed true production reports map to
this root; the diagnostic records contain additional manifestations.
The recorded CPython~3.16-dev native runs show two effects. With the JIT
disabled, lookup returns \path{"after"}; with the JIT enabled, it raises
\path{KeyError}. The control store retains one entry, whereas the JIT store
creates a second entry for the same key object. Both modes were recorded as
not reproduced on 3.14.8 and 3.15.0 after target executors were installed.
The runtime impact is incorrect dictionary lookup and update. Preserving the
rewrite requires the embedded hash to remain consistent with the key's current
type and hash slot.

\subsubsection{Case Study 2: Attribute Acquisition After Finalization}
\label{sec:case-attribute-lifetime}

Late attribute acquisition is the failure behind the running example. The
reference acquires the selected class attribute before releasing the receiver.
The target releases the receiver first, allowing its finalizer to replace the
class attribute and release the old payload before acquisition. This order
occurs in both the with-values and no-dictionary attribute layouts; the
optimized sequence is \path{POP_TOP; _LOAD_CONST_INLINE}.

Figure~\ref{fig:case-attribute-flow} connects the emitted operations to their
runtime effects. The lookup supplies a borrowed pointer $p$; embedding that
pointer does not acquire ownership. Releasing the receiver can invoke its
finalizer, which replaces the supporting class slot and releases $p$ before
the emitted load acquires it.
C1 binds the lookup through the receiver's type to $p$ and preserves the order
of the emitted operations. C2 follows receiver consumption and checks whether
ownership of $p$ was acquired first. The recorded
\path{acquire_before_release} diagnostic identifies the release before that
acquisition. In the lifetime model, the finalizer path loses interval support
and then fails the point requirement for acquisition. This root appears in the
diagnostic-comparator records and contributes no reports to the RQ1
production-precision count.

\begin{figure}[!htbp]
\centering
\begin{minipage}[t]{0.75\linewidth}
\vspace{0pt}
\begin{lstlisting}[style=casecode]
// Optimizer: mortal-attribute branch
p = _PyType_Lookup(type, name);
ADD_OP(_POP_TOP, 0, 0);
ADD_OP(_LOAD_CONST_INLINE, 0, (uintptr_t)p);

// Runtime effects in emitted order
PyStackRef_CLOSE(receiver);
result = PyStackRef_FromPyObjectNew(p);
\end{lstlisting}
\end{minipage}
\caption{Code excerpts for late attribute acquisition.
The CPython~3.16-dev excerpts expand the mortal-attribute branch and the two
emitted operations; names are shortened and unrelated paths omitted.}
\Description{Code excerpts show a borrowed attribute lookup, emission of
receiver release before inline acquisition, and the corresponding runtime
calls.}
\label{fig:case-attribute-flow}
\end{figure}

The recorded native tests isolate the Tier-2 rewrite using builds that correct
the Tier-1 acquisition order. Their JIT-disabled controls also disable
specialization. On CPython~3.15.0 and 3.16-dev, both layout controls return the
old payload, while JIT-active runs read a replacement allocation tagged
\path{junk-0}. This records stale-storage use after finalization and allocator
reuse. The 3.14.8 control fails, leaving that version inconclusive. The
required order is to acquire the attribute before releasing the receiver,
preserving support through the callback.

\subsubsection{Evidence Status}

The catalog contains 15 new project bugs and two public defects. The 92
manually reviewed true production reports map to six roots, identified by
nonzero report counts in Table~\ref{tab:bug-evidence-ledger}. Other roots are
represented by diagnostic-comparator records or separate source checks. The
missing descriptor guard is a recorded miss; stale trace-exit state uses a
separate buffer-state check; and \path{PYTHON-157875} comes from trace-rewind
source analysis~\cite{cpython-issue-157875}.

\begin{table*}[!htbp]
\centering
\footnotesize
\caption{Evidence recorded for the 17 root causes. P denotes a production root
assignment; D denotes a diagnostic-comparator record; Buffer and Source denote
separate source checks. An asterisk marks additional missed manifestations.
Recorded native outcomes are R (recorded reproduced), N (recorded not
reproduced), I (inconclusive), or A (inapplicable). Reports count manually
reviewed true production reports.}
\label{tab:bug-evidence-ledger}
\setlength{\tabcolsep}{2.5pt}
\begin{tabular*}{\linewidth}{@{\extracolsep{\fill}}lrrrrrr@{}}
\toprule
\makecell{Root cause or public issue} &
\makecell{Check path} &
\makecell{Reports} &
\makecell{3.14.8} &
\makecell{3.15.0} &
\makecell{3.16-dev} &
\makecell{Origin} \\
\midrule
Omitted isinstance callback & P, D   & 40 & N & R & R & Project \\
Omitted container callbacks & D      & 0  & I & R & R & Project \\
Stale dictionary hash & P, D*  & 4  & N & N & R & Project \\
Stale initializer code & D      & 0  & I & R & R & Project \\
Zero version in super guard & D      & 0  & N & R & R & Project \\
Stale iterator slots & P, D*  & 2  & N & R & R & Project \\
Zero version in method guard & P, D   & 4  & N & R & R & Project \\
Missing list release & D      & 0  & R & A & A & Project \\
Stale type result & P, D   & 2  & N & R & R & Project \\
Stale mapping slots & D*     & 0  & N & R & R & Project \\
Missing descriptor guard & Missed & 0  & R & R & R & Project \\
Stale isinstance result & P, D   & 40 & N & R & R & Project \\
Stale super lookup & D      & 0  & N & R & R & Project \\
Late attribute acquisition & D      & 0  & I & R & R & Project \\
Stale trace-exit state & Buffer & 0  & N & R & R & Project \\
PYTHON-157875 & Source & 0 & N & R & R & Public \\
PYTHON-158804 & D      & 0 & N & R & R & Public \\
\midrule
Total R & -- & -- & 2  & 15 & 16 & -- \\
Total N & -- & -- & 12 & 1  & 0  & -- \\
Total I & -- & -- & 3  & 0  & 0  & -- \\
Total A & -- & -- & 0  & 1  & 1  & -- \\
\bottomrule
\end{tabular*}
\end{table*}

The version-level records contain reproduced outcomes for two roots on
CPython~3.14.8, 15 on 3.15.0, and 16 on 3.16-dev. Raw execution logs for the
15 project roots are incomplete, so these results are treated as recorded
native outcomes rather than independently revalidated evidence. This
distinction is separate from the manually reviewed production labels.
The project cases participated in development, and their discovery chronology
is incomplete. RQ4 therefore uses the catalog to characterize failure classes,
root causes, and runtime effects. RQ1 provides the controlled accuracy and
frozen-transfer evidence.

\rqfinding{The 17-root catalog spans five failure classes: eight
specialization-dependency, three ownership/lifetime, and two each of
dynamic-effect omission, guard-precondition, and state-protocol failures.
Their recorded effects include stale-storage reads, incorrect lookup results,
retained objects, changed callback behavior, and debug aborts. The cases
exercise source dependencies, temporal lifetime checks, and exit obligations,
showing that rewrite failures extend beyond simple use-after-free errors.}

\section{Discussion}\label{sec:discussion}

\subsection{Implications for Rewrite Correctness}

Lifetime preservation is both temporal and relational. A rewrite can preserve
the value eventually returned while changing whether that value remains
supported at an intermediate program point. RQ3 makes this distinction
concrete: endpoint-only borrow checking misses four violations because support
can disappear and later return. Lifetime preservation therefore requires both
point checks at individual uses and interval checks over the lifetime of a
borrow.
Caller-visible ownership adds a second dimension. A target can satisfy its
local lifetime obligations yet return ownership different from that required
by the caller. Recovering the reference's output role exposes such failures
even when the target has no locally invalid use. Rewrite correctness therefore
depends on both intermediate lifetime and the ownership transferred across the
rewrite boundary.

The reference serves two roles. First, it supplies caller obligations not
explicit in the target: RQ3 shows that recovering an output role exposes one
additional violation and exit requirements expose two more. Second, reference
matching checks whether a locally valid target execution has a safe reference
execution with compatible aliases, supporting owners, exits, and visible
effects. On the selected controls, full matching changes semantic coverage
without adding a report. Reference analysis therefore contributes both
caller obligations and relational coverage.

\subsection{Semantic Coverage and Trust}

A completed check requires sufficient semantics across all three design
stages. Source recovery must identify and bind the relevant actions, lifetime
checking must model effects that establish or invalidate their obligations,
and reference comparison must recover enough behavior to construct a matching
reference execution. Missing information leaves the affected check incomplete.
This distinction matters in production. Of 801 feasible CPython rewrites,
{\tech} reports 194 violations. Among the remaining 607, 43 complete and 564
remain incomplete. Most incomplete cases involve missing source or runtime
bindings, unresolved control-flow or stack-handoff semantics, helper effects,
or incomplete reference witnesses. No recorded reference-coverage query ends
with an undecided solver result. Missing modeled semantics therefore dominate
production incompleteness.

Semantic coverage and report count measure different properties. Adding a
missing contract can complete a previously incomplete check without changing
the report count. Refining an imprecise contract can instead remove a false
report or expose a violation. Reporting completion separately from reports
therefore reveals how much of the extracted population the available semantics
can decide.
Reports depend on the recovered source and runtime specification $\Gamma$.
The analysis uses recovered actions and their order, caller bindings and
aliases, runtime contracts for callbacks and mutations, and reference bindings
for caller-visible behavior. An effect absent from both source recovery and
the runtime model cannot contribute to the checked obligations.

Certificate replay reconstructs the recorded source, bindings, execution path,
state transitions, and solver obligations. It checks that a reported witness
reproduces under the recorded model. Source inspection and native execution
provide separate evidence connecting that model to concrete runtime behavior.

\subsection{Generality and Practical Use}

{\tech}'s lifetime kernel is parameterized by runtime-specific bindings. The
controlled GLib experiment reuses the same lifetime kernel and source lowering
used for CPython while replacing the binding catalog. CPython reference
counting and GLib weak-reference promotion use different ownership protocols,
yet the same checker distinguishes the defective and safe variants in both
controlled populations.
Applying {\tech} to another runtime requires recovery of its compiler
rewrites, contracts for its ownership and callback effects, and a
correspondence between reference and target inputs and exits. Existing
lifetime primitives can be reused when their state transitions match the new
runtime; a new ownership mechanism requires a transfer rule consistent with
the lifetime state model.

For compiler maintainers, each report records the source actions, release
order, supporting owner, failed obligation, path conditions, and relevant
reference information. These witnesses guide source inspection and targeted
runtime tests. Incomplete checks also identify the source or runtime semantics
needed to complete the analysis.

\subsection{Limitations}\label{sec:limitations}

\noindent\textbf{Accuracy and semantic coverage.}
Production review confirms 92 of 194 reports, giving 47.42\% precision.
Production recall is unmeasured because the non-reporting population is not
exhaustively labeled. The 58-input controlled benchmark was constructed and
labeled independently of {\tech} output and later used during implementation.
Its role is controlled accuracy measurement under explicit semantics.
Additional authored, protocol, and source-semantic controls measure mechanism
coverage. The frozen revision studies contain three native-labeled regressions,
all missed because the frozen model lacks required source, caller, or callback
semantics. These cases characterize transfer and model coverage.

\noindent\textbf{Model assumptions.}
The CPython experiments assume valid caller and frame state, execution under
the GIL, and no concurrent external heap mutation beyond effects represented
by the contracts. Runtime callbacks, mutations, allocation outcomes, and
ownership transfers must be represented by the supplied contracts. The
guarantee in Section~\ref{sec:conditional-correctness} applies when source
recovery and runtime bindings satisfy the realization conditions in
Appendix~\ref{app:semantic-interface}. Effects absent from both recovery and
the model remain outside the checked semantics. Replay uses the same transition
semantics as analysis.

\noindent\textbf{Generality and scale.}
Production evaluation covers CPython. The GLib controls demonstrate reuse of
the analysis architecture with a second ownership protocol; production-scale
evaluation on another runtime remains open. The scaling experiments show that
long traces increase solver work and independent branches increase both path
exploration and reference-witness search. The evaluated CPython snapshots
require no explicit path, state, solver, or time cutoff. Larger rewrites and
richer runtime models may require pruning or compositional reasoning.

\noindent\textbf{Engineering effort and validation.}
Two reviewers independently assessed the broader set of 286 production
reports. The final labels classify 92 as true and 194 as false; the default
configuration retains 92 true and 102 false reports. The retained review
record does not contain the initial per-item judgments or review time, so
initial disagreement, inter-rater agreement, and review-effort measurements
are unavailable.
The evaluation measures analysis time and memory but not the effort required
to construct and maintain runtime contracts or review reports. Native outcomes
are recorded for the bug catalog, but raw execution records are incomplete for
the 15 project roots. Modeling effort, report-review effort, and independent
runtime-validation effort are therefore outside the reported analysis cost.

\section{Related Work}\label{sec:related}

\noindent\textbf{Translation validation and rewrite checking.}
Translation validation checks whether an optimized program refines its
reference under a specified semantics~\cite{pnueli-translation-validation}.
Alive2 validates bounded refinement of LLVM transformations using formal IR
semantics~\cite{alive2}. FuzzJIT activates JIT compilation and checks for
execution discrepancies~\cite{fuzzjit}. Optimuzz generates inputs that trigger
selected optimizations and then applies translation validation to those
optimizations~\cite{optimuzz}. These approaches assume that the compared
programs and their semantics are already available, or generate executions
that expose compiler behavior. {\tech} instead reconstructs the comparison
from CPython optimizer source: it recovers the selected reference and target
computations, their path premises and receiving context, and binds their
operations to runtime lifetime contracts. It then checks preservation of
lifetime and caller-visible ownership across the recovered rewrite.

\noindent\textbf{Ownership and lifetime analysis.}
Several systems make ownership explicit in an intermediate representation.
Swift Ownership SSA assigns ownership kinds and constrains their uses in
SIL~\cite{swift-ossa}. CinderX HIR distinguishes owned, borrowed, and uncounted
values, tracks their supporting locations, and models final decrements that can
invoke destructor code and invalidate borrowed
support~\cite{cinderx-refcount}. Perceus derives reference counting and reuse
from a linear resource calculus~\cite{perceus}. {\tech} uses the same
distinction between ownership and borrowed support, but applies it to rewrites
whose lifetime structure is not explicit in an ownership IR. It recovers
ownership-relevant actions from optimizer source and checks point, interval,
and exit obligations across the reference, target, and receiving code.

\noindent\textbf{Python runtime and native-code analysis.}
Pungi, PyRefcon, and CpyChecker detect reference-count and ownership errors in
Python native code using affine analysis, lifecycle modeling, and API ownership
rules~\cite{pungi,pyrefcon,cpychecker-docs}. PyRTFuzz instead uses fuzzing to
exercise the Python interpreter and runtime libraries~\cite{pyrtfuzz}.
{\tech} targets compiler rewrites: it reconstructs reference and target
operations, models their lifetime obligations, and compares caller-visible
effects. Our controlled evaluation directly compares {\tech} with PyRefcon and
CpyChecker on shared C inputs.

\noindent\textbf{Dependencies and runtime pointer protocols.}
Lifetime can depend on state beyond a pointer's immediate owner. CONCURDEP
recovers owner--subject--storage dependencies and checks their uses across
reentrant or parallel events in CPython~\cite{concurdep}. {\tech} uses related
dependency information within rewrite comparison: recovered guards, watcher
registrations, and runtime predicates identify the state that justifies a
cached or borrowed value, and lifetime checking determines whether that state
remains valid when the target uses it.

Runtime interfaces such as LLVM statepoints, HotSpot object maps, JNI handles,
and V8 handle scopes specify pointer requirements at compiled or native
boundaries~\cite{llvm-statepoints,hotspot-oopmap-jdk25,jni-design-java25,v8-embed-handles}.
{\tech} also checks the sequence of effects leading to the boundary. A borrowed
CPython pointer, for example, may initially depend on container ownership and
acquire its own reference only later; the supporting ownership must remain
valid throughout the intervening operations.

\noindent\textbf{Empirical studies of compiler and runtime defects.}
Prior empirical studies classify the symptoms, causes, and repairs of compiler
and JavaScript-engine defects~\cite{sun-compiler-bugs,wang-js-engine-bugs}.
{\tech} complements these taxonomies with an analysis of lifetime failures in
JIT rewrites. Its bug catalog relates observed failures to point, interval,
and exit obligations, while \PyJITBench{} provides controlled inputs for
evaluating the corresponding analysis mechanisms.

\section{Conclusion}\label{sec:conclude}

Compiler rewrites must preserve not only returned values and reference counts,
but also continuous support for borrowed objects across intervening effects and
the ownership required by the caller.
We presented \tech{}, which reconstructs reference and target computations from
optimizer source, binds their runtime effects, and checks point, interval, and
exit obligations. Reference analysis recovers caller obligations and checks
agreement across the modeled execution domain.
On the independently constructed and labeled controlled benchmark, later used
during development, \tech{} detects all 29 defects and reports none of the 29
safe inputs, compared with 20 defects detected by PyRefcon and 18 by
CpyChecker. Across three CPython snapshots, it reports 194 violations among
801 feasible rewrites; manual review classifies 92 as true, giving 47.42\%
precision. Frozen transfer studies expose source and runtime semantics missing
from the fixed model. Overall, the results show that checking temporal support,
caller-visible ownership, and reference agreement exposes rewrite errors that
value and reference-count preservation alone miss. \tech{} identifies 15 new
bugs across five failure classes, each with a reproducible proof of concept.

\ifreferences
  \bibliographystyle{ACM-Reference-Format}
  \bibliography{reference}
\fi

\inlineevidencetrue
\appendix

\section{Semantic and Realization Details}\label{app:semantic-interface}
\subsection{Executable Runtime Contracts}

A fact declaration pairs a predicate with the locations that determine it,
called its \emph{read footprint}. For operands $\bar x$,
\[
 F(\bar x)=\langle\psi_F(\bar x),\mathsf{reads}_F(\bar x)\rangle.
\]
Locations include ownership sets, container slots, type fields, and dependency
tokens. Footprints also include address dependencies: a predicate over
$D[p]$ depends on both $p$ and the selected cell. An interval requirement
additionally records the captured object, its support operands, and the borrow
boundaries.

An action specifies its precondition, guarded outcomes, and read/write
footprints:
\[
 A(\bar x)=
 \langle
   \mathsf{pre},
   \{(g_j,E_j,Q_j,k_j)\}_j,
   R_A,W_A
 \rangle.
\]
Guard $g_j$ selects outcome $j$, $E_j$ gives its ordered effects, $Q_j$
relates the result and poststate to the prestate, and $k_j$ specifies the
control transfer. The footprints $R_A$ and $W_A$ include helper and callback
effects; a \emph{frame condition} identifies locations left unchanged.
The outcomes cover all admitted results, effects, and exits. A callback
without a narrower contract may modify all reachable state. A call that can
affect active borrow support must expose the intermediate change or establish
support throughout its execution. Synchronization clauses restrict
environmental effects between steps.

A successful guard establishes only facts entailed on that branch. A
protection contract relates the protected subject, its registration,
invalidating events, and exit from the specialized path. The required fact
must remain valid until use, or execution must leave the specialized path
before that use. Invalidating a future entry does not protect an execution
already in progress. Each dependency must also correspond to the subject and
registration path of the use it protects. Facts remain distinct: acquiring
ownership establishes liveness but does not establish type or slot stability
and may invalidate uniqueness.

Borrow transfer preserves the active lifetime interval. An exported borrow
carries its captured object, support bindings, path condition, and prior checks
through a named \emph{port}. The receiver imports the port once and establishes
object and support correspondence before executing receiving assignments.
A terminal export records the caller obligation; a supplied continuation
checks later uses and consumption. Transfer preserves the existing interval
and its accumulated checks.

Contracts are encoded as typed JSON clauses over Booleans, unbounded integers,
and integer-indexed arrays. Expressions include equality, Boolean connectives,
linear arithmetic, array reads, and conditionals. Actions contain guarded
outcomes, ordered updates, visible events, exits, and borrow operations. The
loader checks helper schemas and effect consistency, rejects untyped entry
assumptions, ill-typed clauses, and unsupported expressions, and derives write
footprints from action bodies. It also checks declared read footprints against
predicate dependencies, including address dependencies. Unbound variables or
unknown footprints leave the affected action opaque. Certificates identify the
loaded source and specification versions by hash.

\subsection{Concrete Realization}

The conditional result in Section~\ref{sec:conditional-correctness} concerns
the recovered programs under $\Sigma$. Applying it to a selected compiler
rewrite requires source recovery to represent the actions and interfaces
relevant to the checked obligations. The following conditions connect the
symbolic executions to concrete runtime executions.

For side $X\in\{R,T\}$, let $\mathcal A_X(c,a)$ relate concrete state $c$ to
symbolic state $a$. The relation includes control position, object identity,
ownership, storage, and visible events. A binding connects symbolic checks to
concrete executions when it satisfies four obligations.

\begin{enumerate}[leftmargin=*,nosep]
\item \textbf{Entry coverage.}
Every admitted concrete entry pair related by $\rho$ has corresponding
symbolic entries satisfying $P_0$ and the input correspondence.

\item \textbf{Target-step coverage.}
Every concrete target step has a represented outcome that preserves
$\mathcal A_T$ and its visible events. The outcome covers the effects permitted
by $\Gamma$ and either exposes intermediate support changes or establishes
support throughout the call.

\item \textbf{Reference realization.}
Each selected symbolic reference outcome must be realizable from any matching
concrete state. For outcome $j$ with guard $g_j$ and transition $f_j$,
\[
 \mathcal A_R(c,a)\land g_j(a)
 \Longrightarrow
 \exists c'.
 c\xrightarrow{j}c'
 \land
 \mathcal A_R(c',f_j(a)).
\]
The concrete transition also realizes the outcome's declared visible events.

\item \textbf{Predicate and exit agreement.}
A discharged symbolic predicate implies the corresponding concrete point,
interval, or exit obligation. Borrow transfer preserves the interval, and
related symbolic exits imply the required concrete output, alias, support, and
visible-effect relation.
\end{enumerate}
Source-derived bindings establish these obligations by composing primitive
clauses; external bindings state them as trusted contracts. Missing clauses
leave dependent checks incomplete. Reference witnesses additionally require
reference realization for the selected symbolic outcome.
Helper correspondence relates concrete reference outcomes $H_R$ to target
outcomes $H_T$. Under a precondition established at both call sites, a
relation $K$ requires
\[
 \forall y_T\in H_T,\qquad
 \exists y_R\in H_R:\ K(y_R,y_T).
\]
An outcome includes its result, poststate, and visible effects. For
deterministic calls, equal arguments and related state over the complete read
footprint may justify equal results; this state includes ownership changes
that can affect finalization. Other calls require an explicit relational
contract, including null outcomes and fresh-object renaming. Nondeterministic
outcomes without such a contract remain independent. The kernel uses the
outcomes named by the specification; other relational clauses remain
unresolved.
Exit matching compares the caller-visible state carried across the boundary,
including returned objects, ownership, aliases, visible effects, dependencies,
and exported borrows. Exported borrows must agree on their ports, fact schemas,
captured operands, support locations, and support state. The composed receiver
then checks its requirements using the transferred state and accumulated
lifetime history.

\section{Recovery and Runtime Bindings}\label{app:source-recovery}

\subsection{Source Recovery and Evidence Records}
\label{sec:source-details}\label{sec:cpython-instantiation}\label{sec:lifetime-ir}

Source recovery preserves the distinctions needed to reconstruct lifetime and
protection obligations:

\begin{itemize}[leftmargin=*,nosep]
\item \textbf{Abstract locations.}
Ownership-sensitive fields and fields used by indirect calls retain separate
locations and value identities; other fields and array elements use summary
locations. Unknown call targets remain associated with their source sites.

\item \textbf{Ordered rewrites.}
Operation-buffer writes are reconstructed in execution order, including
multiple slots modified along the same optimizer path. Each recovered pair
retains its reference and target slices, operand definitions, and selector
condition. Identical pairs share an identifier and combine selector conditions
by disjunction.

\item \textbf{Incomplete recovery.}
Optimizer proposals retain their source locations. Unsupported writes,
unmatched branches, unsummarized cycles, and values without recovered runtime
producers remain explicit and leave the affected construction incomplete.

\item \textbf{Revision comparisons.}
The repaired definition is the reference and its direct parent is the target,
with a common recovered interface and no optimizer selector condition.
\end{itemize}

Stack composition reconstructs the ownership state passed through the recovered
region:

\begin{itemize}[leftmargin=*,nosep]
\item \textbf{Handle versions.}
Each handle is represented in static single assignment form~\cite{cytron-ssa}. Joins merge
presence and ownership without introducing an acquisition or release.

\item \textbf{Ownership movement.}
Stack assignments copy ownership tags, while transfers remain distinct from
acquisitions. Frame-local reads alias the persistent slot owner; stores
transfer the new owner while retaining cleanup responsibility for the
displaced value.

\item \textbf{Receiving layouts.}
Normal, error, restart, and completed-instruction exits retain distinct stack
layouts. A slot is omitted only when its entry contract and every producer
establish that it carries no ownership.
\end{itemize}

CPython's tagged \path{_PyStackRef} separates object identity from the
ownership state of a particular stack handle
~\cite{cpython-stackref-315,cpython-stackrefs-rq1-doc}. Consuming one handle
therefore need not end the object's lifetime: another owner, runtime root, or
immortality may keep it live. Container-item bindings retain both the captured
item and its supporting container. Thread and frame context pointers are not
treated as returned object handles. Trace-recording-only bodies are excluded
according to the Tier-2 generator, while their stack interfaces and exclusion
evidence remain recorded.
The coverage record $\Lambda$ makes missing semantics explicit:

\begin{itemize}[leftmargin=*,nosep]
\item It links feasible source actions and outcomes to operands, contracts,
requirements, and locations.
\item A site is excluded only when it cannot affect the checked property or
any predicate on which that property depends, including branch conditions.
\item Requirements merged across paths retain their originating sites and path
conditions.
\end{itemize}

Missing indirect targets, unsupported control flow, and incomplete effects
therefore remain visible rather than contributing to a completed check.
Execution records retain the path condition, action and outcome choices,
symbolic store, active borrows, completed requirements, and exit state.
Unsupported relations and undecided solver queries remain unresolved.
Certificates bind these records to source, implementation, bindings,
specification, and provenance hashes needed to reconstruct a reported
execution. Stable expression ordering makes records comparable across
output-directory names; the artifact provides the full schema, manifests,
and hash inventories.

\subsection{CPython and GLib Bindings}
\label{app:cpython-bindings}\label{sec:source-obligations}

Table~\ref{tab:source-rule-specs} summarizes the CPython source patterns for
which {\tech} constructs explicit bindings. Sites outside these categories
remain in the source inventory. Preconditions apply even when an operation's
result is unused, and output roles are indexed by reference path, exit, and
proved entry domain. Unmodeled callbacks, allocation or error paths, and
unresolved storage relations leave dependent checks incomplete.

\begin{table*}[!tb]
  \centering
  \footnotesize
  \caption{CPython source bindings. Dependency bindings use the runtime
  specification's invalidation contract.}
  \label{tab:source-rule-specs}
  \setlength{\tabcolsep}{3pt}
  \begin{tabular}{@{}p{0.17\textwidth}p{0.34\textwidth}p{0.43\textwidth}@{}}
    \toprule
    \textbf{Source category} &
    \textbf{Recovered operands} &
    \textbf{Binding procedure} \\
    \midrule
    Dispatch stability &
    Receiver of a removed iterator or mapping-slot test &
    Match receiver guards, watcher registration, and executor dependencies by
    type and path. \\

    Embedded dispatch &
    Iterator function pointer and its receiver &
    Relate the pointer to its dispatch slot and preceding receiver-type guard. \\

    Embedded value &
    Identity-hash receiver or cached initializer &
    Bind the receiver's type and hash slot; attach the initializer contract
    when available. \\

    Type-version stability &
    Receiver and version operand in a no-op selection &
    Match the selected value to the runtime receiver/version guard. \\

    Receiver-type stability &
    Receiver of an embedded-type or super-descriptor lookup &
    Follow operand definitions and stack movement to the lookup and its
    preceding guard. \\

    Callback preservation &
    Operation replaced by a constant &
    Bind the removed helper's visible effects to its arguments and state
    footprint. \\

    Receiving-stack validity &
    Outputs at restart and completed-instruction exits &
    Construct exit-specific layouts from branch conditions, consumed handles,
    and stack synchronization. \\

    Support until acquisition &
    Descriptor, owning slot, receiver, and consuming actions &
    Follow the borrowed operand to acquisition; exclude paths that exit first
    or release unrelated inputs. \\

    Guard precondition &
    Version-tag operand of an emitted guard &
    Relate the operand at use to its selection-time value and required nonzero
    condition. \\
    \bottomrule
  \end{tabular}
\end{table*}

The bindings fall into three groups:

\begin{itemize}[leftmargin=*,nosep]
\item \textbf{Lifetime bindings.}
These connect a captured object to the ownership that supports it until use.
Descriptor bindings relate the payload, owning slot, lookup type, receiver,
and consuming action. Path conditions distinguish a release followed by use
from releases on exiting paths or releases of unrelated objects. Final-release
bindings expand deallocator callbacks only after the updated reference count
permits finalization. Receiving-stack bindings preserve ownership transfer and
use the exit-specific layout, including restart synchronization.

\item \textbf{Specialization bindings.}
These connect cached state to the guards and dependencies that protect it.
The selected receiver, successful type or version guard, watcher registration,
and executor dependency must identify the same protected subject. The
invalidation contract must preserve the required field until use or force
execution off the specialized path. Cached hashes depend on both the
receiver's type and hash slot. Replaced calls retain visible effects unless
their arguments, read state, and path conditions justify removing them.
Unsummarized paths or intervening effects without a protection proof remain
unresolved.

\item \textbf{GLib weak-reference bindings.}
A weak reference permits its referent to disappear, whereas a borrow requires
continuous support. \path{g_weak_ref_get} atomically acquires a strong
reference or returns null when the weak reference is empty
~\cite{glib-weakref-get}. A rewrite that separates the pointer read from the
acquisition must therefore protect against final release between those steps.
The GLib controls replace the CPython-specific promotion and weak-reference
bindings while reusing the source lowerer and lifetime checker. Their native
oracle covers a sequential object lifecycle; concurrent transitions are
represented by the runtime contract.
\end{itemize}

\section{Evaluation Details}\label{app:population-registry}
\subsection{Benchmark Populations and Configurations}
\label{app:controlled-fixtures}

Table~\ref{tab:evidence-roles} separates the roles of the evaluation
populations. The shared 58-input benchmark was constructed and labeled
independently of {\tech} output and later used during implementation. Its case
selection and expected outcomes were fixed independently of checker reports.
The 12 C-API cases encode six defects and six changed-safe variants under
declared ownership conditions. The remaining 46 inputs encode 23 historical
defect/repair mechanisms, with each repaired encoding used as an identity
control. The artifact binds each historical encoding to vulnerable and repair
revisions, a source location, a repair rationale, and source hashes. All three
tools receive the same candidate C inputs; {\tech} additionally receives the
reference and runtime contracts.

\begin{table*}[!htbp]
\centering\footnotesize
\caption{Evidence roles, benchmark provenance, and development exposure.}
\label{tab:evidence-roles}
\setlength{\tabcolsep}{3pt}
\begin{tabularx}{\linewidth}{@{}p{0.17\linewidth}Xp{0.20\linewidth}p{0.18\linewidth}@{}}
\toprule
\textbf{Population} & \textbf{Construction and exposure} &
\textbf{Labels} & \textbf{Evidence role} \\
\midrule
58 portable C inputs &
Independently constructed and labeled; subsequently used during development &
29 defect, 29 safe & Controlled accuracy and baseline comparison \\
Authored C / JIT-shaped &
Constructed and used during development &
Defect/safe pairs & Mechanism breadth \\
Protocol / source controls &
Developed to exercise components and runtime semantics &
Registered expectations; selected native oracles & Component validation \\
Production CPython &
Three pinned snapshots; receiving-context refinement used this population &
Reviewed reports; non-reports not exhaustively labeled & Precision and semantic coverage \\
Frozen revisions &
Selection before prediction; fixed checker and contracts &
Three native defects; separate preservation labels & Transfer and model-coverage gaps \\
\bottomrule
\end{tabularx}
\end{table*}

Table~\ref{tab:benchmark-selection} gives the selection rules and label
sources. The label types have different meanings:

\begin{itemize}[leftmargin=*,nosep]
\item \textbf{Defect/safe} labels state the expected behavior under a given
input contract.
\item \textbf{Preservation} labels relate two implementations; both may contain
the same defect.
\item \textbf{Registered expectations} specify internal analysis outcomes and
may intentionally require an incomplete check.
\item \textbf{Identity} controls compare an implementation with itself, while
\textbf{changed-safe} controls compare different implementations satisfying
the same contract.
\end{itemize}

\begin{table*}[!htbp]
\centering\footnotesize
\caption{Benchmark selection and label provenance. Units differ across rows.
The 18 source-semantic comparisons and four inputs for each runtime form the
26-comparison source-pipeline study.}
\label{tab:benchmark-selection}
\setlength{\tabcolsep}{3pt}
\begin{tabularx}{\linewidth}{@{}p{0.18\linewidth}p{0.10\linewidth}XXl@{}}
\toprule
\textbf{Population} & \textbf{Units} & \textbf{Selection} &
\textbf{Oracle / label} & \textbf{RQ} \\
\midrule
Shared portable C &
58 inputs &
12 modified C-API inputs and 46 encodings of 23 historical mechanisms &
Independently specified defect/repair labels; six changed-safe and 23 repaired
identity controls &
1, 2 \\

Authored C &
32 inputs &
Development pairs covering ownership patterns &
Defect/changed-safe labels &
1 \\

JIT-shaped &
12 inputs &
Development pairs using the rewrite interface &
Defect/safe labels &
1 \\

Protocol controls &
46 inputs &
Registered kernel and incomplete-model variants &
Internal-state expectations, not independent defect labels &
1, 3 \\

Source semantics &
18 comparisons &
Nine source/producer, five descriptor, and four final-release variants &
Registered expectations; native observations on declared schedules &
1--3 \\

CPython bindings &
4 inputs &
Acquisition-order variants &
Source contracts and acquisition/finalization oracles &
3 \\

GLib bindings &
4 inputs &
Promotion, early-unlock, and omitted-acquisition variants &
Binding contracts; separate sequential native lifecycle check &
3 \\

Production &
3,332 alternatives &
All extracted rewrites in three pinned snapshots &
Manual review of reported witnesses; non-reports lack exhaustive labels &
1, 2 \\

Scaling &
18 pairs &
Generated identity pairs varying actions, branches, outcomes, or borrows &
Registered structure and internal-state expectations &
2 \\

Changed functions &
10 pairs &
Every eligible common function in a fixed source interval &
One native defect; other pairs lack independent safety labels &
1 \\

Historical window &
105 pairs &
Next eight registered history positions &
96 generated-C preservation labels; nine unlabeled &
1 \\

Beta window &
13 pairs &
Every eligible common definition in a fixed release interval &
Native, source-only, preservation, and unlabeled cases &
1 \\

Bug catalog &
17 roots &
15 project findings and two public defects &
Static records and recorded native status, kept separate &
4 \\
\bottomrule
\end{tabularx}
\end{table*}

Native controls provide concrete oracles for the runtime contracts:

\begin{itemize}[leftmargin=*,nosep]
\item \textbf{Descriptor and final-release controls.}
The schedules vary mutation, receiver/type aliasing, shared payloads,
replacement aliasing, and registration. The callback installs a replacement,
invalidates the dependency, and releases the old payload. Monitors check
identity, ownership, support, and event order before invalid access. Five
descriptor variants use 32 schedules each; four final-release variants use
64 each.

\item \textbf{Source-result controls.}
The list-lookup oracle distinguishes a non-null slot, a null slot, and an
out-of-range lookup with an error indicator. Ownership applies to the object
actually returned. The nullable-allocation control instead uses a declared
two-outcome machine and does not require both outcomes to occur natively.

\item \textbf{GLib controls.}
The native oracle checks identity after promotion, liveness after releasing
the caller's reference, finalization on the last strong release, and null after
the weak reference is cleared. It covers one sequential lifecycle rather than
concurrent schedules.
\end{itemize}

The artifact retains the complete source variants and native schedule
inventories.

\noindent\textbf{Configurations.}\label{app:tool-configurations}
The portable baselines receive the same C inputs; {\tech} additionally
receives the reference computation and runtime contracts. PyRefcon omits two
obsolete code-constructor models for header compatibility. CpyChecker uses
GCC~6.5.0 with Python~3.6 and records per-function exceptions as errors.
For component studies, the \emph{full} diagnostic configuration enables source
checks, paired checks, and shared-warning cancellation. The
\emph{local + feasible effects} configuration preserves the same graphs,
premises, helper outcomes, feasible paths, and ordered effects, but removes
warning cancellation and paired output, alias, guard-domain, and exit-kind
checks. Both configurations use the same source rules and, where applicable,
the same nullable-owned output contract. Protocol-kernel results remain
separate from diagnostic-comparator outcomes.

\subsection{Frozen Transfer and Model-Coverage Selection}

Table~\ref{tab:frozen-selection} records the three populations selected before
prediction. Each pair is checked in forward, reverse, and identity directions:
parent/changed, changed/parent, and changed/changed. The frozen records bind
the checker, contracts, source bodies, labels, and prediction commands. They
also retain additions, removals, and ineligible commits. These studies test
transfer of the fixed model to separately selected revisions. Their three
native defect labels support case-level analysis of the observed misses.

\begin{table*}[!htbp]
\centering\footnotesize
\caption{Frozen revision selection and pre-prediction evidence. Short Git
identifiers and release tags locate the exact intervals; full manifests are
in the artifact.}
\label{tab:frozen-selection}
\setlength{\tabcolsep}{3pt}
\begin{tabularx}{\linewidth}{@{}p{0.17\linewidth}XXr@{}}
\toprule
\textbf{Study} & \textbf{Selection rule} &
\textbf{Evidence fixed before prediction} & \textbf{Pairs} \\
\midrule
Changed functions &
All commits touching \path{Objects/} after \texttt{0bb9ef9695d3} through
\texttt{be3dfe5b1890}; changed C functions present on both sides &
One dictionary regression labeled by native control/failure trials; other
pairs have no independent safety label &
10 \\

Historical window &
Positions 201--208 in the bytecode history at \texttt{0bb9ef9695d3}; four
commits supply changed common definitions &
Identical generated interpreter and Tier-2 C gives preservation labels to
96 pairs from one syntax edit; nine unlabeled &
105 \\

Beta window &
All non-merge commits in \path{v3.13.0b1..v3.13.0rc1} touching bytecodes or
three optimizer files; nine commits supply eligible definitions &
Two native regressions, one source-only monitoring-event expectation, five
preservation labels, and five unlabeled cases &
13 \\
\bottomrule
\end{tabularx}
\end{table*}

The selection details are:

\begin{itemize}[leftmargin=*,nosep]
\item \textbf{Changed functions.}
Selection preceded prediction, although commit subjects and two
allocation-failure diffs had been inspected. None of the twenty selected
function bodies matches the retained three-snapshot production bodies. The
native dictionary oracle distinguishes a present entry with a null value from
an absent entry.

\item \textbf{Historical window.}
The study registers positions 201--208 after positions 1--200 had been
inspected, without filtering by subject or predicted outcome. Its 96
preservation labels come from one mechanical syntax change and represent one
change mechanism across multiple pairs.

\item \textbf{Beta window.}
Selection covers the full beta interval after an empty maintenance interval
and an initial five-definition beta interval had been considered. Those first
five definitions had been inspected and remain in the final population. The
two native labels concern a missing-key callback and iterator exhaustion; both
are callback/state regressions.
\end{itemize}

The later-snapshot reuse run and earlier development revision studies remain
in the artifact as reuse and development-exposure studies.

\subsection{Production Coverage Diagnostics}

Table~\ref{tab:unknown-causes} summarizes the 564 incomplete non-reporting
checks in the canonical production configuration. Categories overlap: 428
inputs have multiple gaps and 47 have one. The remaining 89 have no missing
modeled behavior on executed paths but lack complete reference-witness
coverage.

\begin{table}[!htbp]
\centering\small
\caption{Coverage gaps in the canonical population of 564 incomplete checks.
Counts overlap; the last row has no missing modeled behavior on executed paths
but still lacks a complete reference witness.}
\label{tab:unknown-causes}
\begin{tabularx}{\linewidth}{@{}Xr@{}}
\toprule
\textbf{Missing semantics or reference coverage} & \textbf{Inputs} \\
\midrule
Operand, storage, or expression binding & 452 \\
Control flow or stack handoff & 254 \\
Helper effect contract & 189 \\
Assertion binding & 165 \\
Destructor behavior & 171 \\
Helper or macro binding & 130 \\
Partial helper outcomes & 2 \\
No unmodeled behavior; reference witness incomplete & 89 \\
\bottomrule
\end{tabularx}
\end{table}

\noindent
Two earlier repair experiments on the broader population show how missing
semantics affect completion:

\begin{itemize}[leftmargin=*,nosep]
\item \textbf{Restart-input binding.}
Following the Tier-2 stack flush removes the corresponding missing-binding
diagnostic from 87 inputs but leaves the report and completion counts
unchanged.

\item \textbf{Multiline-condition parsing.}
Preserving Boolean precedence completes four additional load/pop checks,
raising completed non-reports from 39 to 43 while leaving the broader
286-report set unchanged. All 43 completed checks remain in the canonical
configuration.
\end{itemize}

The artifact retains the before/after manifests and per-input category
assignments.

\section{Evidence and Audit Details}\label{app:production-audits}
\subsection{Secondary Analysis Audits}

Separate C++ requirement-checker and diagnostic-comparator runs support model
development and source auditing. They analyze different units and are excluded
from the RQ1 production-precision population. Their role is diagnostic:

\begin{itemize}[leftmargin=*,nosep]
\item \textbf{Model gaps.}
The auxiliary runs identify missing receiving contexts, incomplete bindings,
and source-model corrections.

\item \textbf{Grouping audits.}
We compare grouping by selector and operation with grouping that also preserves
path conditions, source bodies, interfaces, operand bindings, and stack
conditions. Review labels are not used as grouping keys. A group containing
different labels exposes distinctions lost by the grouping rule; a
label-homogeneous group provides no direct measure of review effort.

\item \textbf{Witness audit.}
An earlier automated audit identified caller-condition concerns in ten accepted
reports and lacked a report-to-cause connection for some other accepted
labels. The subsequent manual review reassessed all production reports against
their witnesses and source context. The final classification contains 92 true
and 194 false reports in the broader configuration. RQ1 uses the canonical
subset, which retains all 92 true reports and 102 false reports.
The retained record contains the final classifications but not the reviewers'
initial per-item judgments, so the initial disagreement rate cannot be
reconstructed.
\end{itemize}

The artifact retains the auxiliary per-input outputs, group memberships,
inspection samples, false-report categories, automated findings, manual-review
record, and diagnostic inspection plot.

\subsection{Bug-Evidence Provenance}
\label{app:bug-chronology}

RQ4 combines three evidence sources:

\begin{itemize}[leftmargin=*,nosep]
\item \textbf{Static evidence} records report labels and diagnostic root
assignments.
\item \textbf{Native evidence} records whether a root is reproduced under a
specified build, execution mode, control, and failure oracle.
\item \textbf{External evidence} records public issue reports and upstream
responses separately from project findings.
\end{itemize}

Revalidating a native label requires matching source and binary versions,
proof-of-concept revision, build configuration, environment, evidence that the
target code executed, and a cause-specific control/failure oracle. A successful
control establishes the expected behavior; executor installation alone does
not establish the reported failure.
For the 15 project roots, 72 version/mode rows refer to 144 unavailable
control/failure logs. The repository audit found no same-named plain or
compressed copies in the searched locations. RQ4 therefore reports these
results as author-recorded native outcomes, separately from the manually
reviewed static labels.
on CPython~3.14.8, the current report records a failing control and an
inconclusive outcome, whereas an older native table records non-reproduction
with passing exits. The evidence ledger follows the current report and records
the discrepancy for revalidation.
The corresponding 3.15 and development-build observations also depend on
specific controls. The native-JIT control repairs the Tier-1 acquisition order
while retaining the Tier-2 rewrite. The JIT-disabled control additionally
disables specialization because disabling native code generation alone leaves
specialization active. The stale-storage outcomes are therefore recorded with
these control conditions.

\end{document}